\documentclass{aa}
\input epsf.tex
\begin{document}

\thesaurus{08.23.2; 08.09.2; 08.02.2; 08.13.2; 08.22.3; 08.03.4}
\title{Abundance correlations in mildly metal-poor stars\thanks{Based on 
       observations obtained at the European Southern Observatory, La Silla, 
       Chile.}}
\author{E. Jehin, 
        P. Magain\thanks{Ma\^{\i}tre de Recherches au Fonds National de la 
                         Recherche Scientifique (Belgium)}, 
        C. Neuforge\thanks{Charg\'e de Recherches au Fonds National de la 
                         Recherche Scientifique (Belgium)}, 
        A. Noels, 
        G. Parmentier and 
        A. A. Thoul} \offprints{E.\ Jehin}
\institute{Institut d'Astrophysique et de G\'eophysique, Universit\'e de 
           Li\`ege, 5, Avenue de Cointe, B-4000 Li\`ege, Belgium}
\date{Received date; accepted date}
\maketitle
\markboth{E.\ Jehin et al.: 
          Abundance correlations in mildly metal-poor stars}{}

\begin{abstract}
Accurate relative abundances have been obtained for a sample of 21 mildly 
metal-poor stars from the analysis of high resolution and high signal-to-noise
spectra.  In order to reach the highest coherence and internal precision, 
lines with similar dependency on the stellar atmospheric parameters were 
selected, and the analysis was carried out
in a strictly differential way within the sample.

With these accurate results, correlations between relative abundances have 
been searched for, with a special emphasis on the neutron capture elements.

This analysis shows that the $r$ elements are closely correlated to the 
$\alpha$ elements, which is in agreement with the generally accepted idea 
that the $r$-process takes place during the explosion of massive stars.

The situation is more complex as far as the $s$ elements are concerned.  Their 
relation with the $\alpha$ elements is not linear.  In a first group of stars,
 the relative abundance of the $s$ elements increases only slightly with the 
$\alpha$ elements overabundance until the latter reaches a maximum value.  
For the second group, the $s$ elements show a rather large range of 
enhancement and a constant (and maximum) value of the $\alpha$ elements 
overabundance.

This peculiar behaviour leads us to distinguish between two sub-populations of
metal-poor stars, namely Pop$\,$IIa (first group) and Pop$\,$IIb 
(second group).

We suggest a scenario of formation of metal-poor stars based on two distinct 
phases of chemical enrichment, a first phase essentially consisting in 
supernova explosions of massive stars, and a second phase where the enrichment
is provided by stellar winds from intermediate mass stars.
More specifically, we assume that all thick disk and field halo stars were born
in globular clusters, from which they escaped, either during an early 
disruption of the cluster (Pop$\,$IIa) or, later, through an evaporation 
process (Pop$\,$IIb).


\end{abstract}

\section{Introduction}

The chemical composition of the solar system is now known with a high accuracy 
from the spectroscopic studies of the solar photosphere as well as from the 
analysis of the carbonaceous chondrites (Anders and Grevesse \cite{AG},
Grevesse et al. \cite{NOELS}).  
Moreover, theoretical models of stellar evolution and
nucleosynthesis allow the identification of the most likely sites and 
mechanisms for the formation of the chemical elements. Combining them with 
simple models for the chemical evolution of the Galaxy, one can reproduce 
reasonably well the solar system abundances.

However, the solar system only provides a single data point, both in the time 
evolution of the Galaxy and in the spatial distribution in the disk.  
It is thus important to gather additional data in order to have an insight 
of the chemical composition of the interstellar matter at different epochs 
since the formation of the Galaxy.  

As time goes by, more and more metals are synthesized in the stellar interiors
and ejected in the surrounding environment, so that the overall metallicity 
slowly increases. Although the correlation
of metallicity with time is not a perfect one, unevolved metal-poor stars 
give a picture of the chemical conditions at earlier phases of the galactic 
evolution.

Up to now, spectroscopic analyses of metal-poor stars have essentially 
consisted in determining abundance ratios as a function of metallicity.  
These ratios are supposed to show the evolution of the nucleosynthetic 
processes with time.  Unfortunately, these results
generally show a rather large scatter, which is either of cosmic origin or 
due to observational uncertainties.  In the latter case, this prevents any 
meaningful comparison with the theoretical models.  On the other hand, if 
the scatter reflects an intrinsic
cosmic dispersion, it should be accounted for in the models.

In order to find the origin of that scatter, we must improve the accuracy 
of the abundances. This requires both high quality data and a careful 
spectroscopic analysis.

Here, we report on the results of such a detailed analysis and we
investigate the underlying reasons for the remaining scatter -- if any.  We
look at correlations  between different abundance ratios at a given 
metallicity.
Strongly correlated elements are likely to have been processed in the same 
astrophysical
sites.  Therefore, the identification of such correlations can provide 
fruitful insights on the nucleosynthesis of these elements.  This allows us 
to suggest a 
scenario for the formation of metal-poor stars.  Preliminary results have 
already been presented briefly in Jehin et al.\  (1998).

\section{Determination of the atmospheric parameters}

\begin{table}
\caption[]{Basic photometric data}
\begin{flushleft}
\begin{tabular}{|rrllll|}
\hline
   ~HD  &ID &~V    &b$-$y &V$-$K     & ~$c_1$     \\
\hline                     	      
22879  &1  &6.70  &0.366 &1.478	&0.274  \\
25704  &2  &8.12  &0.371 &1.53 	&0.275  \\
59984  &3  &5.92  &0.355 &1.423	&0.335  \\
61902  &4  &8.24  &0.329 &1.29 	&0.361  \\
63077  &5  &5.38  &0.372 &1.53 	&0.273  \\
63598  &6  &7.94  &0.366 & --  	&0.299  \\
76932  &7  &5.83  &0.360 &1.475	&0.299  \\
78747  &8  &7.73  &0.383 & --  	&0.296  \\
79601  &9  &8.00  &0.378 & -- 	&0.306  \\
97320  &10 &8.19  &0.337 &1.31 	&0.305  \\
111971 &11 &8.05  &0.353 & --  	&0.304  \\
126793 &12 &8.19  &0.373 &1.44 	&0.291  \\
134169 &13 &7.68  &0.368 &1.464	&0.309  \\
152924 &14 &8.02  &0.318 &1.24 	&0.379  \\
189558 &15 &7.74  &0.385 &1.575	&0.269  \\
193901 &16 &8.67  &0.376 &1.52 	&0.221  \\
194598 &17 &8.35  &0.343 &1.38 	&0.269  \\
196892 &18 &8.24  &0.346 &1.37 	&0.307  \\
199289 &19 &8.29  &0.368 &1.42 	&0.264  \\
203608 &20 &4.22  &0.326 &1.23 	&0.310  \\
215257 &21 &7.42  &0.357 &1.53 	&0.310  \\
\hline
\end{tabular}
\end{flushleft}
\end{table}

\begin{table}
\caption[]{Model parameters}
\begin{flushleft}
\begin{tabular}{|ccccc|}
\hline
   HD     &$T_{\rm eff}$  &$\log g$  &[Fe/H]  &$\xi$  \\
          &(K)        &(cgs)  &        &(km/s)  \\
\hline                                 
22879     &5774     &4.1    &$-1.00$   &1.0  \\  
25704     &5715     &3.9    &$-1.05$   &1.1  \\
59984     &5860     &3.8    &$-0.85$   &1.3  \\
61902     &6054     &3.9    &$-0.85$   &1.3  \\
63077     &5717     &4.0    &$-0.90$   &0.9  \\ 
63598     &5779     &3.9    &$-0.95$   &1.1  \\
76932     &5798     &3.9    &$-1.00$   &1.0  \\ 
78747     &5676     &3.8    &$-0.80$   &1.0  \\
79601     &5716     &3.8    &$-0.75$   &1.1  \\
97320     &6000     &4.1    &$-1.30$   &1.2  \\
111971    &5880     &4.0    &$-0.85$   &1.1  \\
126793    &5783     &3.9    &$-0.90$   &1.3  \\
134169    &5782     &3.8    &$-0.90$   &1.2  \\
152924    &6132     &3.9    &$-0.80$   &1.4  \\
189558    &5628     &3.8    &$-1.20$   &1.2  \\
193901    &5703     &4.4    &$-1.15$   &0.8  \\
194598    &5920     &4.2    &$-1.20$   &1.0  \\
196892    &5925     &4.0    &$-1.10$   &1.2  \\
199289    &5808     &4.1    &$-1.15$   &1.2  \\
203608    &6112     &4.3    &$-0.80$   &1.3  \\
215257    &5769     &3.8    &$-0.90$   &1.3  \\
\hline
\end{tabular}
\end{flushleft}
\end{table}

\section{Observations and data reduction}

We have selected a sample of 21 unevolved metal-poor stars (Tables 1 and 2), 
with roughly one tenth of the
solar metallicity ([Fe/H] $\sim -1$ \footnote {We adopt the usual 
spectroscopic 
notation: [A/B]  $\equiv \log _{10} (N_A / N_B)_{\ast} - 
\log _{10} (N_A / N_B)_{\odot}$ for elements A and B.} ).  
This corresponds more or less to the transition between the halo and the disk.

The observations were carried out with the Coud\'e Echelle Spectrometer 
(CES) fed by the 1.4\,m Coud\'e Auxilliary Telescope (CAT) at the European 
Southern Observatory (La Silla, Chile). The long camera was used with a Loral 
CCD detector (ESO \#38) having 2688$\times$512 pixels of 
15$\times$15 $\mu$m each.

The exposure times were chosen in order to reach a signal-to-noise ratio 
(S/N) of at least 200 in all spectral regions.
The spectra were collected during two observing runs, in January 1996 
(8 nights) and July 1996 (11 nights), the last one having been carried out 
in remote control from the ESO headquarters in Garching bei M\"unchen. 
All stars were observed in four 
wavelength bands, each band having a width of about 50\AA . These spectra are 
centered around 4125, 4325, 4584 and 4897\AA.

The data reduction consisted in :\\[-3ex]
\begin{enumerate}
\item[1.]background subtraction on the basis of the mean level measured on 
the parts of the CCD not illuminated by stellar light;
\item[2.]flat-fielding, using the spectrum of an internal lamp;
\item[3.]wavelength calibration, using the stellar lines themselves to 
define the 
calibration curve, thus automatically correcting for the radial velocity;
\item[4.]definition of the continuum, in the form of a low order Spline fitted 
through a number of pre-defined continuum windows;
\item[5.]equivalent widths (EWs) measurements, by Gaussian and Voigt function 
fitting, the first method being preferred for the weak lines and the second 
in the case of the stronger ones (for which the Lorentzian damping wings 
contribute 
significantly to the EW). The table listing the nearly 2000 EWs 
measurements is available in electronic form from CDS.
\end{enumerate}

\subsection{Effective temperature}

The effective temperatures $T_{\rm eff}$ were determined from the 
Str\"omgren b$-$y and Johnson V$-$K colour indices, using the calibration 
of Magain (\cite {M87}) which is based on the infrared flux method (Blackwell
and Shallis \cite{BS}). The sources of b$-$y measurements are 
Schuster and Nissen (\cite {SN}), Olsen (\cite {OL}) and Carney 
(\cite {CA83}). The V$-$K colours were 
obtained from Carney (\cite{CA83}), Alonso et al.\  (\cite {AL}) and from some 
unpublished
measurements by P. Magain with the ESO 1\,m telescope on La Silla. 
When several values were available for the same star, the average was taken. 

The adopted effective temperature is a mean of the two determinations and 
is listed in Table 2. The agreement between the temperatures deduced from 
the two colour 
indices is quite good, the mean difference amounting to 45\,K, with an r.m.s. 
scatter of 40\,K in the individual determinations.

All these stars being in the solar neighbourhood (distances range from 
9 to 80 pc), the interstellar reddening should be negligible.  The internal 
precision of the 
effective temperatures may thus be estimated from the comparison of the two 
photometric determinations.
The scatter of 40\,K in the $T_{\rm eff}$ differences corresponds to an 
uncertainty of 28\,K
($40 / \sqrt{2}$) in the individual values and of 20\,K in the mean 
$T_{\rm eff}$ from the two colour indices.  This high internal precision 
can be checked from the differential
excitation equilibria of Fe{\sc i}.  The excitation equilibria indicate 
effective temperatures
which agree quite well with the photometric values:  the scatter of the 
excitation temperatures
around the photometric ones amounts to 57\,K, while the expected scatter, on 
the basis of a precision of 20\,K on the photometric values and 45\,K on 
the excitation values, amounts to 50\,K.  The 7\,K difference is completely 
negligible and shows that the internal precision of
our effective temperatures is indeed very high.

\subsection{Metallicity}

For the first step, the model metallicities were taken from previously 
published analyses. This is not a crucial parameter in metal-poor stars model 
atmospheres as the continuous opacity is dominated by the contribution of the 
negative hydrogen ion
and hydrogen itself is the main electron donor.  Nevertheless, the model
metallicities were redetermined on the basis of our set of Fe{\sc i} lines.
The latter values were used in the subsequent analysis.

\subsection{Surface gravity}

Surface gravities are usually determined by requiring  Fe{\sc i} and 
Fe{\sc ii} lines to indicate the same abundance. However, this procedure is 
affected by several uncertainties. For example, the iron abundance derived 
from Fe{\sc i}
lines is quite sensitive to $T_{\rm eff}$ and unfortunately the zero-point 
of the $T_{\rm eff}$ scale of metal-poor stars may be uncertain by as much 
as 100\,K, causing errors in [Fe/H] of about 0.10 dex and errors in the 
spectroscopic gravities of the order of 0.25 dex. For other uncertainties, 
see for e.g the discussion in Nissen et al.\  (1998).
To avoid these problems, we decided (1) to determine the surface gravity
from the Str\"omgren $c_1$ index, using the calibrations of 
VandenBerg and Bell (\cite {VB}) for the adopted temperatures and 
metallicities; (2) to always derive
relative abundances from the comparison of lines of the same ionization 
stage and, thus, of the same dependence on surface gravity. 
 
\subsection{Microturbulence velocity}

Microturbulence velocities $\xi$ are obtained, as usual, by forcing a set of 
lines of the same element and same ionization stage but with different EWs 
to indicate the same abundance. In our observations, only the Fe{\sc i} 
lines are suitable for such a 
determination.  The precision of the microturbulence determinations, 
estimated by linear regression, is better than 0.1 km s$^{-1}$.
The adopted model parameters for the 21 stars are listed in Table 2.

\section{Method of analysis}

\subsection{A strictly differential analysis}
Our stellar sample is rather homogeneous in terms of atmospheric parameters:
all the stars are dwarfs or subgiants with $3.8 < \log g < 4.4$, have 
roughly solar temperatures with $5620\,{\rm K} < T_{\rm eff} < 6140\,{\rm K}$ 
and a narrow range of metallicity  ($-1.3 <$ [Fe/H] $< -0.7$). A differential 
analysis within the sample is thus indicated, especially as we are interested 
in distinguishing minute variations from star to star and, so, need the highest
possible internal precision.

As a first step, all stars were analysed with respect to HD\,76932, one of the 
brightest stars in our sample, having average atmospheric parameters.
The zero points of the element abundances in HD\,76932 were determined from
lines with laboratory $gf$-values (Table 3).  When lines from two ionization 
stages were available,  the abundance of the ion was forced to agree with 
that of the neutral.

The microturbulence velocity in the atmosphere of HD\,76932, which must 
be known in order for this differential analysis to be carried out, 
was determined from
a set of Ca{\sc i} lines with the same excitation potential, precise laboratory
$gf$-values (Smith and Raggett \cite{SR81}) and a suitable range of EWs 
(Table 3). These lines were observed with 
the same instrument in the context of another programme.  A value of 
$\xi = 1.0$ km/s has been obtained.

It is possible to reduce even further the uncertainties in the relative 
$gf$-values
by analysing each star with respect to each other and derive mean $gf$-values
from the whole sample.  This reduces the scatter in the line abundances for an
individual star but has no effect on the mean abundances of one star relative
to another, provided that exactly the same lines are used in both stars.  Since
this is not the case for all stars, we only used this global differential 
analysis whenever justified, i.e.\  when the number of lines of a given 
species is large enough and when the line sample varies from star to star.
Such an analysis was performed for Fe{\sc i}, Fe{\sc ii}, Ti{\sc i}, 
Ti{\sc ii}, Cr{\sc i}, Cr{\sc ii} and Ni{\sc i}.

\begin{table}
\caption[]{Data for the 2.52 eV Ca{\sc i} lines in HD$\,$76932: 
wavelength $\lambda$,
excitation potential $\chi$, oscillator strength $\log gf$, equivalent 
width EW and
damping enhancement factor $f_6$ over the Uns\"old formula.}
\begin{flushleft}
\begin{tabular}{cccccc}
\hline
 $\lambda$  &$\chi$ &$\log gf$ &EW     &$f_6$  \\
 (\AA)       &(eV)  &          &(m\AA) &       \\
\hline
5260.390   &2.52  &$-1.719$  &\ 8.6 &1.4  \\
5261.710   &2.52  &$-0.579$  &54.2   &1.4  \\
6161.295   &2.52  &$-1.266$  &23.4   &2.0  \\
6163.754   &2.52  &$-1.286$  &25.3   &2.0  \\
6166.440   &2.52  &$-1.142$  &29.1   &2.0  \\
6169.044   &2.52  &$-0.797$  &49.9   &2.0  \\
6169.564   &2.52  &$-0.478$  &66.3   &2.0  \\
6493.788   &2.52  &$-0.109$  &83.4   &0.8  \\
6499.654   &2.52  &$-0.818$  &47.9   &0.8  \\
\hline
\end{tabular}
\end{flushleft}
\end{table}

\begin{table}
\begin{flushleft}
\caption[4a.]{Atomic line data\\
The six columns give, respectively, the element and its ionization stage, 
the wavelength, the excitation potential of the lower level, the differential $\log gf$
determined from our analysis, the absolute $\log gf$ from the literature
and the EW in the spectrum of HD\,76932.}
\begin{tabular}{|cccrrr|}
\hline
El.  &$\lambda$ &$\chi$  &$\log gf$   &$\log gf$      &HD$\,$76932 \\
     &(\AA)     &(eV)    &dif.      &abs.         &(m\AA)  \\
\hline 
\hline                                
Mg\,I  &4571.102  &0.00  &$-$5.014  &$-$5.550     &76.0    \\
       &         &        &          &            &        \\
Ca\,I  &4108.532  &2.71  &$-$0.547   &            &31.9    \\
Ca\,I  &4578.559  &2.52  &$-$0.415   &$-$0.697    &48.5    \\
       &          &       &          &            &        \\
Sc\,II &4314.091  &0.62  & 0.378    &$-$0.040     &98.1    \\
Sc\,II &4320.749  &0.61  & 0.147    &$-$0.210     &87.8    \\
      &            &       &         &            &        \\
Ti\,I  &4112.716  &0.05  &$-$1.648  &$-$1.758     &13.9    \\
Ti\,I  &4870.136  &2.25  & 0.501    &$+$0.358       &16.0    \\
Ti\,I  &4885.088  &1.89  & 0.497    &             &29.0    \\
Ti\,I  &4913.622  &1.87  & 0.267    &$+$0.160       &20.7    \\
       &          &         &        &            &        \\ 
Ti\,II &4316.802  &2.05  &$-$1.498  &     &34.2    \\
Ti\,II &4330.245  &2.05  &$-$1.552  &             &32.7    \\
Ti\,II &4330.708  &1.18  &$-$1.892  &     &53.8    \\
Ti\,II &4563.766  &1.22  &$-$0.599  &     &105.4   \\
Ti\,II &4568.328  &1.22  &$-$2.752  &             &17.6    \\
Ti\,II &4571.982  &1.57  &$-$0.100  &     &124.0   \\
Ti\,II &4583.415  &1.16  &$-$2.739  &     &19.6    \\
Ti\,II &4589.953  &1.24  &$-$1.446  &     &70.4    \\
Ti\,II &4874.014  &3.09  &$-$0.809  &     &22.5    \\
Ti\,II &4911.199  &3.12  &$-$0.493  &     &34.7    \\
       &            &      &         &            &        \\
V\,I   &4111.787  &0.30  &$-$0.047  &$+$0.408       &42.6    \\
V\,I   &4115.177  &0.29  &$-$0.407  &$+$0.071       &28.1    \\
V\,I   &4330.024  &0.00  &$-$1.148  &$-$0.631     &13.7    \\
V\,I   &4577.184  &0.00  &$-$1.618  &$-$1.048     &5.6     \\
V\,I   &4594.126  &0.07  &$-$1.232  &$-$0.672     &10.7    \\
V\,I   &4875.492  &0.04  &$-$1.355  &$-$0.807     &9.3     \\
       &           &       &        &             &        \\
Cr\,I  &4111.358  &2.90  &$-$0.442  &             &11.5    \\
Cr\,I  &4129.184  &2.91  &$+$0.246    &             &35.4    \\
Cr\,I  &4580.062  &0.94  &$-$1.580  &             &38.2:   \\
Cr\,I  &4600.757  &1.00  &$-$1.376  &$-$1.276     &43.5    \\
Cr\,I  &4870.816  &3.08  &$+$0.162    &             &25.8    \\
Cr\,I  &4885.774  &2.54  &$-$2.117  &$-$1.055     &5.5     \\
       &        &         &          &            &        \\
Cr\,II &4558.650  &4.07  &$-$0.465  &             &55.5    \\
Cr\,II &4588.204  &4.07  &$-$0.686  &             &47.2    \\
Cr\,II &4592.057  &4.07  &$-$1.293  &             &23.5    \\
Cr\,II &4884.598  &3.86  &$-$2.083  &             &9.0     \\
       &          &      &           &            &        \\
Fe\,I  &4109.062  &3.29  &$-$1.464  &             &36.0    \\        
Fe\,I  &4112.323  &3.40  &$-$1.701  &             &20.1    \\     
Fe\,I  &4114.451  &2.83  &$-$1.366  &             &56.5    \\    
Fe\,I  &4114.942  &3.37  &$-$1.644  &             &22.8    \\  
Fe\,I  &4120.212  &2.99  &$-$1.290  &             &53.8    \\
Fe\,I  &4124.489  &3.64  &$-$2.219  &             &4.7     \\
Fe\,I  &4125.886  &2.84  &$-$2.056  &             &28.3    \\
Fe\,I  &4126.191  &3.33  &$-$0.963  &             &57.5    \\
Fe\,I  &4126.857  &2.84  &$-$2.763  &             &7.8     \\
Fe\,I  &4132.908  &2.84  &$-$0.997  &$-$0.960     &71.0    \\
Fe\,I  &4136.527  &3.37  &$-$1.573  &             &25.1    \\
Fe\,I  &4137.005  &3.41  &$-$0.679  &             &62.2    \\
\hline
\end{tabular}
\end{flushleft}
\end{table}

\begin{table}\setcounter{table}{3}
\begin{flushleft}
\caption[4b.]{Atomic line data (contd.)}
\begin{tabular}{|cccrrr|}
\hline
El.  &$\lambda$ &$\chi$  &$\log gf$   &$\log gf$      &HD$\,$76932 \\
     &(\AA)     &(eV)    &dif.      &abs.         &(m\AA)  \\
\hline 
\hline   
Fe\,I  &4137.415  &4.28  &$-$0.961  &             &17.4    \\
Fe\,I  &4566.524  &3.30  &$-$2.228  &             &10.7:   \\
Fe\,I  &4574.225  &3.21  &$-$2.419  &             &8.4     \\
Fe\,I  &4574.728  &2.28  &$-$2.889  &             &18.3    \\
Fe\,I  &4587.134  &3.57  &$-$1.727  &             &14.7    \\
Fe\,I  &4595.365  &3.30  &$-$1.712  &             &26.6    \\
Fe\,I  &4596.416  &3.65  &$-$2.169  &             &6.0     \\
Fe\,I  &4598.125  &3.28  &$-$1.529  &             &34.2    \\
Fe\,I  &4602.008  &1.61  &$-$3.182  &$-$3.154     &30.2    \\
Fe\,I  &4602.949  &1.48  &$-$2.195  &$-$2.220     &76.0    \\
Fe\,I  &4875.881  &3.33  &$-$1.900  &             &18.2    \\     
Fe\,I  &4885.434  &3.88  &$-$1.045  &             &30.3    \\   
Fe\,I  &4886.337  &4.15  &$-$0.735  &             &33.4    \\ 
Fe\,I  &4896.442  &3.88  &$-$1.931  &             &6.2     \\
Fe\,I  &4907.735  &3.43  &$-$1.804  &             &18.3    \\
Fe\,I  &4908.032  &4.22  &$-$1.562  &             &7.4     \\
Fe\,I  &4910.020  &3.40  &$-$1.348  &             &38.5    \\
Fe\,I  &4910.330  &4.19  &$-$0.742  &             &30.0    \\
Fe\,I  &4910.570  &4.22  &$-$0.754  &             &31.4    \\
Fe\,I  &4911.782  &3.93  &$-$1.687  &             &9.0     \\
Fe\,I  &4917.235  &4.19  &$-$1.025  &             &19.9    \\
Fe\,I  &4918.015  &4.23  &$-$1.212  &             &14.4:   \\
       &         &       &           &            &        \\
Fe\,II &4128.742  &2.58  &$-$3.668  &     &24.4    \\
Fe\,II &4576.339  &2.84  &$-$3.047  &     &41.0    \\
Fe\,II &4582.833  &2.84  &$-$3.280  &     &30.8    \\
       &          &       &         &             &        \\
Ni\,I  &4331.651  &1.68  &$-$2.146  &$-$2.100     &31.6    \\  
Ni\,I  &4600.364  &3.60  &$-$0.499  &             &24.2    \\ 
Ni\,I  &4604.996  &3.48  &$-$0.252  &$-$0.250     &38.4    \\   
Ni\,I  &4606.226  &3.60  &$-$0.921  &             &11.3:   \\
Ni\,I  &4873.446  &3.70  &$-$0.459  &$-$0.380     &22.1    \\   
Ni\,I  &4904.418  &3.54  &$-$0.079  &             &45.0    \\
Ni\,I  &4912.025  &3.77  &$-$0.739  &             &12.4    \\
Ni\,I  &4913.978  &3.74  &$-$0.569  &             &17.3    \\
Ni\,I  &4918.371  &3.84  &$-$0.140  &             &29.9    \\
       &          &      &          &             &        \\ 
Sr\,I  &4607.338  &0.00  &          & 0.280       &14.9    \\
        &        &         &         &            &        \\
Y\,II  &4883.690  &1.08  &$+$0.074    & 0.070     &42.3    \\
Y\,II  &4900.124  &1.03  &$-$0.072  &$-$0.090     &38.7    \\
       &          &        &         &            &        \\
Zr\,II &4317.321  &0.71  &          &$-$1.380     &8.3     \\
       &         &     &             &            &        \\
Ba\,II &4130.657  &2.72  &           & 0.560      &18.8    \\
       &         &       &           &            &        \\
La\,II &4322.505  &0.17  &$-$0.938  &             &5.4     \\
La\,II &4333.763  &0.17  &$-$0.152  &$-$0.160     &22.4    \\
       &         &        &          &            &        \\
Ce\,II &4137.655  &0.52  &$+$0.065   &            &10.2    \\
Ce\,II &4562.367  &0.48  &$-$0.080   & 0.330      &9.1     \\
       &           &    &            &            &        \\
Nd,II &4109.450  &0.32  &$+$0.548   & 0.519      &20.0    \\
Nd\,II &4314.512  &0.00  &$-$0.292  &$-$0.226     &8.4     \\
      &        &          &          &            &        \\
Sm\,II &4318.936  &0.28  &          &$-$0.270     &6.4     \\
       &         &       &          &             &        \\
Eu\,II &4129.724  &0.00  &          & 0.204       &38.8    \\
\hline
\end{tabular}
\end{flushleft}
\end{table}

\subsection{Model atmospheres and line analysis}

The abundance analysis was carried out with model atmospheres constructed 
individually for each star. The usual assumptions of Local Thermodynamic 
Equilibrium (LTE) 
and plane parallel (horizontally homogeneous) atmospheres were made. 
These models were  calculated on the basis of the temperature stratifications 
($T(\tau)$ relations) of Kurucz (1993).
Given the $T(\tau)$ relation, the gas pressure, the electron pressure and the
continuous absorption were computed with a programme based on
the Gustafsson et al.\  (\cite{GU}) subroutines.
 
When more than one line of the same species was measured for a star, the mean 
abundance value and the standard deviation were computed.
Whenever possible, the analysis was restricted to lines having EWs
between 5 and 50\,m\AA. Using weaker lines would lead to increased random 
errors (and possibly some systematic overestimates), while stronger lines 
are very sensitive to microturbulence and damping. In a few cases, however, 
especially for
Mg, rather strong lines had to be used, while only very weak lines
were available for Zr, Sm and Ce.  The damping constants $\gamma$ were 
computed with the Uns\"old formula (Gray 1972), with an empirical 
enhancement factor of 1.5.

The 93 spectral lines used in the abundance analysis are listed in 
Table 4. They were selected after a careful inspection of the 
stellar and solar spectra. 
The lines were chosen as far as possible to be free of blends affecting the EW
measurements.

\section{Element abundances}

\subsection{Iron peak elements}

We have obtained the abundances of four iron peak elements, namely V, Cr, Fe 
and Ni (Table 5).
They all have a rather large number of clean lines and the line-to-line scatter
of the abundances is small ($<\sigma> \,\sim\, 0.03$). In the case of Fe 
and Cr, both neutral and ionized lines are available.
 
Our iron abundances are based on 34 Fe{\sc i} and 3 Fe{\sc ii} transitions.
The zero point for iron 
was obtained from 3 Fe{\sc i} lines with accurate 
oscillator strengths from the Oxford group (Blackwell et al.\  1982a and 
references therein). 
The iron ionization equilibrium is satisfied to a good accuracy 
($<$[Fe/H]$_{ \rm I}-$[Fe/H]$_{ \rm II}> = +0.011 \pm 0.051$), the small 
scatter confirming the 
photometrically determined surface gravities.

From the 6 Cr{\sc i} and 4 Cr{\sc ii} lines available in our spectra, 
we find that 
the chromium abundance in metal-poor stars scales like the iron one. 
The scatter around the
mean is very small in both cases: 0.030 dex for [Cr/Fe]$_{ \rm I}$ 
and 0.032 dex for [Cr/Fe]$_{ \rm II}$. The agreement between the neutral 
and ionized lines is satisfactory:
$<$[Cr/H]$_{ \rm I}-$[Cr/H]$_{ \rm II}> =$ 
\newline
$+0.002 \pm 0.050$.  
The zero point was fixed
 from the oscillator strengths of Blackwell et al.\  (1984, 1986b) 
for Cr{\sc i}.

Nickel is represented in our spectra by 9 clean Ni{\sc i} lines.  
The three absolute $\log gf$ available for the zero point calibration are from 
Doerr and Kock (\cite{DK85}) and the recent determinations from 
Wickliffe and Lawler (\cite{WL97}).  

Because of the lack of hyperfine structure (HFS) data for our V{\sc i} lines, 
we have restricted our analysis to 6 lines with EWs smaller 
than 30\,m\AA\  (except 4111.787\AA\  which is slightly stronger but had to 
be included for our most metal-poor stars). 
Using accurate transition probabilities from Whaling et al. (\cite{W85}) we 
have derived gf-values for each V{\sc i} line. Absolute abundances were thus 
computed and compared with the results obtained from the differential 
analysis. The very small scatter around the mean difference 
($<\sigma> \,\sim\, 0.02$) reflects the high internal accuracy of the two 
sets of $gf$-values.  The relative abundances
of iron peak elements are listed in Table 5.

\begin{table*}
\caption[]{Iron peak element abundances and line-to-line scatter}
\begin{tabular}{ccccccc}
\hline
 ID  &[Fe/H]$_{ \rm I}$  &[Fe/H]$_{ \rm II}$  &[V/Fe]$_{ \rm I}$ &[Cr/Fe]$_{ \rm I}$  &[Cr/Fe]$_{ \rm II}$ &[Ni/Fe]$_{ \rm I}$  \\
\hline
\hline		             
  1 &$-$0.892(16) &$-$0.888(12) &$-$0.032(36) &$-$0.111(28) &$-$0.107(26) &$-$0.082(32) \\
  2 &$-$0.960(26) &$-$1.001(19) &$\;$   0.015(33) &$-$0.124(33) &$-$0.109(30) &$-$0.093(37) \\
  3 &$-$0.755(27) &$-$0.755(07) &$-$0.103(32) &$-$0.098(43) &$-$0.128(15) &$-$0.112(33) \\
  4 &$-$0.727(23) &$-$0.693(16) &$-$0.102(36) &$-$0.098(51) &$-$0.130(49) &$-$0.122(29) \\
  5 &$-$0.831(22) &$-$0.840(10) &$\;$   0.030(35) &$-$0.079(38) &$-$0.052(56) &$-$0.070(33) \\
  6 &$-$0.856(27) &$-$0.970(25) &$-$0.046(66) &$-$0.101(41) &$-$0.099(41) &$-$0.086(36) \\
  7 &$-$0.910(29) &$-$0.910(12) &$-$0.039(37) &$-$0.089(35) &$-$0.089(22) &$-$0.083(31) \\
  8 &$-$0.730(20) &$-$0.857(11) &$-$0.033(34) &$-$0.098(55) &$-$0.070(17) &$-$0.084(32) \\
  9 &$-$0.668(23) &$-$0.760(17) &$\;$   0.008(29) &$-$0.072(26) &$-$0.023(24) &$-$0.074(31) \\
 10 &$-$1.220(17) &$-$1.179(31) &$-$0.095(82) &$-$0.124(32) &$-$0.108(46) &$-$0.080(35) \\
 11 &$-$0.737(27) &$-$0.748(25) &$-$0.136(30) &$-$0.157(56) &$-$0.105(33) &$-$0.134(30) \\
 12 &$-$0.800(24) &$-$0.864(17) &$-$0.001(55) &$-$0.113(47) &$-$0.117(31) &$-$0.080(30) \\
 13 &$-$0.804(32) &$-$0.823(10) &$-$0.069(37) &$-$0.107(43) &$-$0.086(15) &$-$0.095(46) \\
 14 &$-$0.708(28) &$-$0.675(41) &$-$0.106(140)&$-$0.088(62) &$-$0.102(49) &$-$0.082(32) \\
 15 &$-$1.129(34) &$-$1.093(30) &$-$0.037(38) &$-$0.098(51) &$-$0.132(40) &$-$0.108(38) \\
 16 &$-$1.071(25) &$-$1.043(06) &$-$0.200(58) &$-$0.154(38) &$-$0.143(31) &$-$0.232(44) \\
 17 &$-$1.126(30) &$-$1.074(28) &$-$0.139(57) &$-$0.164(33) &$-$0.141(76) &$-$0.185(45) \\
 18 &$-$1.031(21) &$-$1.025(13) &$-$0.024(81) &$-$0.082(45) &$-$0.064(33) &$-$0.077(37) \\
 19 &$-$1.074(32) &$-$1.100(18) &$-$0.023(63) &$-$0.142(45) &$-$0.065(48) &$-$0.055(39) \\
 20 &$-$0.677(29) &$-$0.636(07) &$-$0.079(51) &$-$0.109(93) &$-$0.133(30) &$-$0.086(37) \\
 21 &$-$0.804(29) &$-$0.814(15) &$-$0.160(90) &$-$0.162(36) &$-$0.088(39) &$-$0.136(43) \\
\hline
\end{tabular}
\end{table*}

\subsection{$\alpha$ elements}

The magnesium abundance is obtained from a single Mg{\sc i} line, the calcium 
abundance from two neutral lines and the titanium abundance from a first set 
of 4 Ti{\sc i} lines and a second set of 8 Ti{\sc ii} lines.

Only the rather strong intercombination MgI line \newline
($\sim$70\,m\AA) 
at 4571\,\AA\   is available in our spectra for the Mg abundance determination.
The deduced abundance is thus  more sensitive to the microturbulence 
velocity $\xi$. 

The line-to-line scatter in the deduced titanium abundances is $\sim 0.025$.
The ionization equilibrium of titanium confirms, once again, the validity
of the photometric surface gravities: 
$<$[Ti/H]$_{ \rm I}-$[Ti/H]$_{ \rm II}> = +0.016 \pm 0.044$.
The zero point was determined  from 3 accurate Ti{\sc i} $gf$-values  
(Blackwell et al.\  1982b, 1986a). 

The relative abundances are given in Table 6.

\begin{table}
\caption[]{$\alpha$ element abundances and line-to-line scatter}
\begin{tabular}{cclcc}
\hline
ID  &[Mg/Fe]\,I  &[Ca/Fe]\,I  &[Ti/Fe]\,I  &[Ti/Fe]\,II  \\
\hline
\hline
  1 &0.344   &0.197(37)  &0.249(20)  &0.234(22) \\
  2 &0.269   &0.133(82)  &0.217(74)  &0.180(46) \\
  3 &0.239   &0.065(39)  &0.105(32)  &0.126(27) \\
  4 &0.077   &0.050(59)  &0.102(32)  &0.077(21) \\
  5 &0.344   &0.175(27)  &0.230(46)  &0.264(22) \\
  6 &0.299   &0.185(37)  &0.260(29)  &0.258(35) \\
  7 &0.366   &0.242      &0.241(31)  &0.241(21) \\
  8 &0.322   &0.260(24)  &0.233(32)  &0.271(38) \\
  9 &*       &0.198(24)  &0.227(25)  &0.251(23) \\
 10 &0.303   &0.169(55)  &0.260(20)  &0.217(57) \\
 11 &0.096   &0.054(47)  &0.058(30)  &0.048(29) \\
 12 &0.341   &0.185(27)  &0.244(28)  &0.263(23) \\
 13 &0.326   &0.152(86)  &0.182(39)  &0.186(14) \\
 14 &0.234   &0.132(88)  &0.173(44)  &0.172(49) \\
 15 &0.398   &0.242(42)  &0.256(62)  &0.234(34) \\
 16 &0.164   &0.119(60)  &0.071(38)  &0.101(34) \\
 17 &0.139   &0.177      &0.120(52)  &0.091(43) \\
 18 &0.324   &0.222(39)  &0.272(29)  &0.251(31) \\
 19 &0.338   &0.167(46)  &0.262(40)  &0.235(20) \\
 20 &0.152 &$-$0.007(60) &0.072(35)  &0.029(23) \\
 21 &0.063   &0.084(33)  &0.033(44)  &0.032(27) \\
\hline
\end{tabular}
\end{table}

\begin{table*}
\caption[]{Heavy element abundances and line-to-line scatter}
\begin{tabular}{crrrrrrrrr}
\hline
ID & [Sr/Fe]I & [Y/Fe]II & [Zr/Fe]II & [Ba/Fe]II & [La/Fe]II & [Ce/Fe]II & [Nd/Fe]II  & [Sm/Fe]II & [Eu/Fe]II \\
\hline
\hline
 1 &$-$0.099  &$-$0.033(13)  & 0.155    & 0.180    &0.241(39)  & 0.007(13)    &$-$0.103(21)  & 0.208    & 0.188  \\ 
 2 &$-$0.119  &$-$0.174(19)  &$-$0.009  & 0.078    &0.141(104) &$-$0.105(31)  &$-$0.055      & 0.170    & 0.161  \\
 3 &$-$0.231  &$-$0.195(09)  &$-$0.020  & 0.048    &0.120      &$-$0.125      &$-$0.190(18)  & 0.017    & 0.080  \\
 4 &$-$0.236  &$-$0.264(32)  &$-$0.027  &$-$0.005  &0.067(26)  &$-$0.165(16)  &$-$0.251      & 0.007    &$-$0.007\\ 
 5 &$-$0.127  &$-$0.070(19)  & 0.084    & 0.092    &0.220(13)  &$-$0.134      &$-$0.114(51)  & 0.164    & 0.195  \\
 6 &$-$0.145  & 0.005(54)    & 0.195    & 0.145    &0.200(80)  &$-$0.104(39)  &$-$0.093(33)  & 0.287    & 0.205  \\
 7 &$-$0.086  & 0.045(15)    & 0.190    & 0.119    &0.198      &$-$0.050      &$-$0.098      & 0.242    & 0.210  \\
 8 &$-$0.170  &$-$0.050(19)  & 0.159    & 0.109    &0.208      &$-$0.064      &$-$0.064(60)  & 0.187    & 0.162  \\
 9 &$-$0.233  &$-$0.151(21)  & 0.054    & 0.055    &0.147      &$-$0.165(63)  &$-$0.073(19)  & 0.218    & 0.185  \\
10 & 0.003    &$-$0.150(34)  &     *    & 0.060    &0.100      &$-$0.189      & 0.046        & 0.267    & 0.229  \\
11 &$-$0.314  &$-$0.218(32)  &$-$0.153  & 0.026    &0.126      &$-$0.074      &$-$0.258(27)  &$-$0.126  &$-$0.027\\
12 &$-$0.118  &$-$0.102(20)  & 0.174    & 0.141    &0.258      &$-$0.067(39)  &$-$0.136(25)  & 0.191    & 0.184  \\
13 &$-$0.276  &$-$0.193(39)  & 0.027    & 0.057    &0.117(61)  &$-$0.104(45)  &$-$0.202(15)  & 0.051    & 0.148  \\
14 &$-$0.279  &$-$0.196(48)  &$-$0.070  &$-$0.086  &0.072(62)  &$-$0.167      &$-$0.293      &$-$0.022  & 0.135  \\
15 & 0.002    & 0.117(31)    & 0.363    & 0.396    &0.338(95)  & 0.112(66)    &$-$0.033(54)  & 0.229    & 0.223  \\
16 &$-$0.383  &$-$0.207(49)  &     *    & 0.147    &0.288      & 0.076        &$-$0.016      & 0.374    & 0.363  \\
17 &$-$0.351  &$-$0.230(31)  &$-$0.106  &$-$0.015  &0.175      &$-$0.023      &$-$0.025      & 0.283    & 0.214  \\
18 &$-$0.103  & 0.048        & 0.353    & 0.155    &0.275(16)  & 0.056(28)    &$-$0.104(25)  & 0.313    & 0.200  \\
19 &$-$0.198  &$-$0.138(35)  & 0.109    & 0.038    &0.276(134) & 0.008(111)   & *            & 0.117    & 0.185  \\
20 &$-$0.254  &$-$0.223(20)  &$-$0.032  & 0.038    &0.135(22)  &$-$0.045(20)  &$-$0.203(07)  & 0.149    & 0.026  \\
21 &$-$0.329  &$-$0.309(16)  &$-$0.095  & 0.054    &0.079(33)  &$-$0.123      &$-$0.286      & 0.016    &-0.006  \\
\hline
\end{tabular}
\end{table*}

\subsection{Heavy elements}

For the sake of clarity, we shall distinguish between the light 
$s$-process elements
(Sr, Y, Zr), the heavy $s$-process elements (Ba, La, Ce) and the mixed or
$r$-process elements (Nd, Sm, Eu).  This is of course an oversimplification, 
several processes contributing in variable proportions to the synthesis 
of each of them.

For Sr, we used the neutral resonance line at 4607.34 \AA\  with an 
oscillator strength $\log gf = 1.92 \pm 0.06$ (Migdalek and Baylis 
(\cite {MB87})), a value which is in agreement with both experimental and 
theoretical determinations.

The agreement between our two lines of Y{\sc ii} is always very good, with a 
mean difference of 0.02 between the two line abundances. 
The absolute $\log gf$ used were determined from measurements of radiative 
lifetimes and branching ratios (Hannaford et al. \cite {HL82}).

The only Zr{\sc ii} line available in our spectra has an accurate 
laboratory oscillator strength (Bi\'emont et al.\  \cite {BG81}).

The only Ba line available is the 4130 \AA\  Ba{\sc ii} line. The 
abundances were computed with 9 HFS components but the HFS effect turns out 
to be very small. 
The transition probability was taken from Gallagher (\cite {G67}). 
This line is slightly blended on the red wing with a Ce{\sc ii} line but the 
error on the deduced differential abundances should not exceed 5\%.

For the La{\sc ii} line at 4333 \AA, we have adopted the transition 
probability calculated by Gratton and Sneden (\cite{GS94}) from the 
experimental lifetime measurements of Arnesen et al.\  (\cite{AB77a, AB77b}) 
and Andersen et al.\  (\cite {AP75}). No HFS was included because de EWs 
never exceed 30 m\AA. This line is slightly blended by a CH line but no 
correction was applied.

Two lines of Ce{\sc ii} are measurable in our spectra but are located in very 
crowded regions. Moreover, these two lines are not available  for all the 
stars. 
For the zero point, we have adopted the transition probability given by 
Gratton and Sneden (\cite{GS94}) for the 4562 \AA\ line, determined by 
renormalizing the values of Corliss and Bozman (\cite{CB62}) to the lifetimes 
results of Andersen et al. (\cite {AP75}). The abundance derived 
for HD\,97320 is very uncertain.

The Nd{\sc ii} lines are also not easy to measure and are not available
for all stars.  We have performed an inverse solar analysis to obtain $\log gf$
values, but the zero point is somewhat uncertain.

The Sm abundance was derived from a single weak Sm{\sc ii} line, and the zero
point deduced from the oscillator strength given by Bi\'emont et 
al.\  (\cite {BG89}).

The Eu{\sc ii} line at 4129.724 \AA\  has both a large HFS and a significant 
isotopic shift. Each of the two isotopes has 16 HFS components
(Woolf et al.\  1995). They were all included, based
on the data from Brostr\"om et al. (\cite {BM95}) for the isotopic shift 
(4578 MHz). 
The 32 HFS components were calculated using data from Becker et al. 
(\cite {BE93}), Villemoes and Wang (\cite {VW94}) and  M\"oller et al. 
(\cite {MH93}). 
A laboratory $gf$-value ($\log gf = 0.204 \pm 0.027$), determined from 
lifetimes and relative line intensities (Bi\'emont et al.\  \cite{B82}) 
was used. The isotopic ratio was assumed to be identical to the meteoritic 
value.
For each star we have computed synthetic spectra between 4129.5 and 
4130.1 \AA\  including also 5 Fe{\sc i} lines with $\log gf$ from Grevesse 
and Sauval (private communication). The macroturbulence velocities 
were determined from a set of 5 clean Fe{\sc i} lines in this spectral window.
The Eu abundance was adjusted until the synthetic and observed spectra 
matched. 

The abundances of the neutron capture elements relative to Fe are given in 
Table 7.

\begin{figure}
\begin{center}
\leavevmode 
\epsfxsize= 8.5 cm 
\epsffile{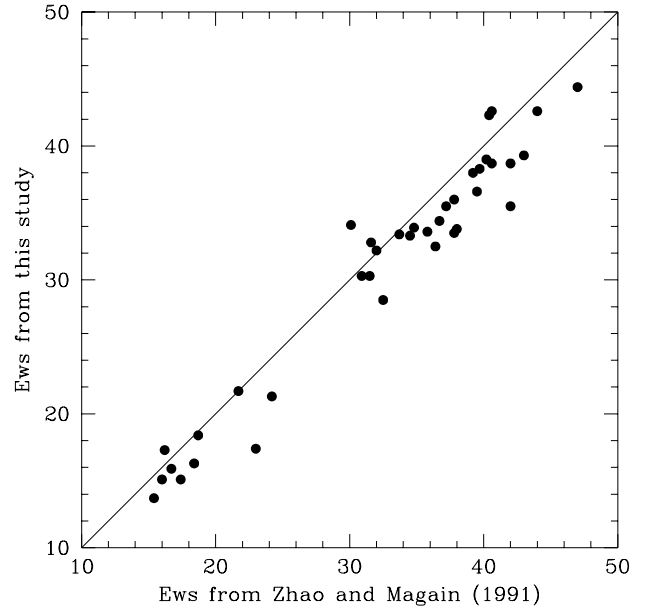}
\caption{Comparison of our EWs with those measured by Zhao and Magain (1991)}
\end{center}
\end{figure}

\begin{figure}
\vspace*{-3.3cm}
\begin{center}
\leavevmode 
\epsfxsize= 8.5 cm 
\epsffile{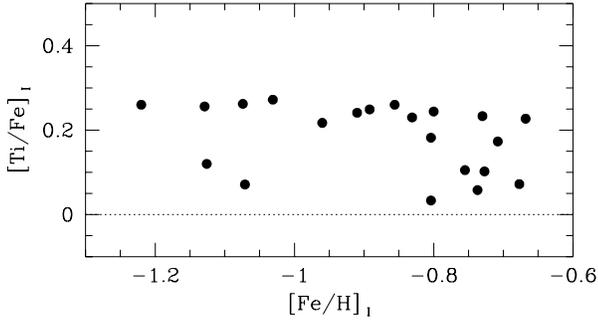}
\vspace*{-5mm}
\caption{Traditional plot: the abundance of Ti relative to Fe versus [Fe/H]}
\end{center}
\end{figure}

\section{Estimation of the uncertainties}

We now discuss the uncertainties on the abundances.  First, we consider the 
errors which act on single lines and, secondly, those which affect whole 
sets of lines.

In the first category, we have the uncertainties on the line parameters, 
i.e., the random errors on EWs, on oscillator strengths, 
on damping constants,...  

They can be estimated by comparing the results from different lines of the 
same element. With 30  Fe{\sc i} lines
 ($5\,$m\AA\  $<$ EW $< 70\,$m\AA), the scatter varies somewhat from star 
to star (from 0.016 to 0.034) with a mean value of 0.027 dex. We find 
similar values for other elements having a significant number of lines 
in our spectra (Ni{\sc i}, Ti{\sc ii},...). 

Our EW measurements for the 10 stars which also appear in 
Zhao and Magain (\cite{ZM91}) are plotted
in Fig.\  1.  The small systematic difference is due to a non-linearity of 
the CCD used by
Zhao and Magain, which accounts for an overestimate of about 5\% 
in their EWs (Gosset and Magain 1993).  Apart from that, 
the agreement is very good, with a scatter  of 1.5\,m\AA\  only. 
Thus, if the two studies were of the same quality, the precision on the EWs 
would be 1\,m\AA. For typical lines having EWs of 25\,m\AA\  the 
corresponding error on the abundance is of the order of 0.02 dex.

Considering these two different estimates (0.027 and 0.02 dex), the 
total abundance error
 coming from EW measurements can be assumed to amount to $0.025/ \sqrt{N}$, 
where $N$ is the number of lines measured.  

The second category of errors are essentially 
model errors, such as uncertainties on the effective temperature, surface 
gravity, microturbulence velocity, overall metallicity and temperature 
structure.
We estimate uncertainties in the differential $T_{\rm eff}$ values due to 
errors in observed photometric indices to be about 20\,K (see Section 3.1). 
In $\log g$ the total errors may approach 0.2 dex, resulting from errors in 
the $c_1$ index, and uncertainties in the calibration of 
VandenBerg and Bell (\cite{VB}). If there is a systematic error in 
$T_{\rm eff}$, it will affect our $\log g$ determination. 
However, the three ionization equilibria available confirm the differential 
photometric gravities (Sections 5.1 and 5.2).  A systematic error on all 
surface gravities will not change the abundance ratios.
Moreover, the comparison with the ``Hipparcos parallax-based gravities" 
of Nissen et al.\  (1998) for the four stars in common shows that, if present,
 such a systematic
error is very small, of the order of $0.12 \pm 0.03$ dex.
 
\begin{figure}
\begin{center}
\leavevmode 
\epsfxsize= 8.5 cm 
\epsffile{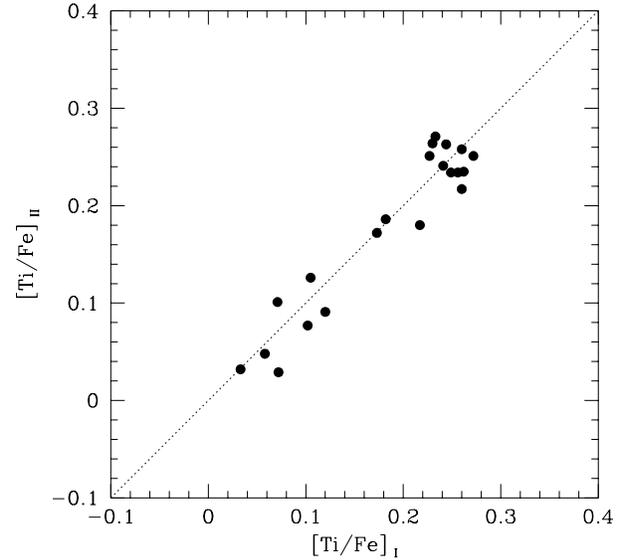}
\vspace*{-5mm}
\caption{Comparison of the values of [Ti/Fe] determined from ionic and from neutral lines}
\end{center}
\end{figure}

\begin{figure*}
\begin{minipage}{8.5cm}
\begin{center}
\leavevmode 
\epsfxsize=8.5cm
\epsfysize=8.5cm 
\epsffile{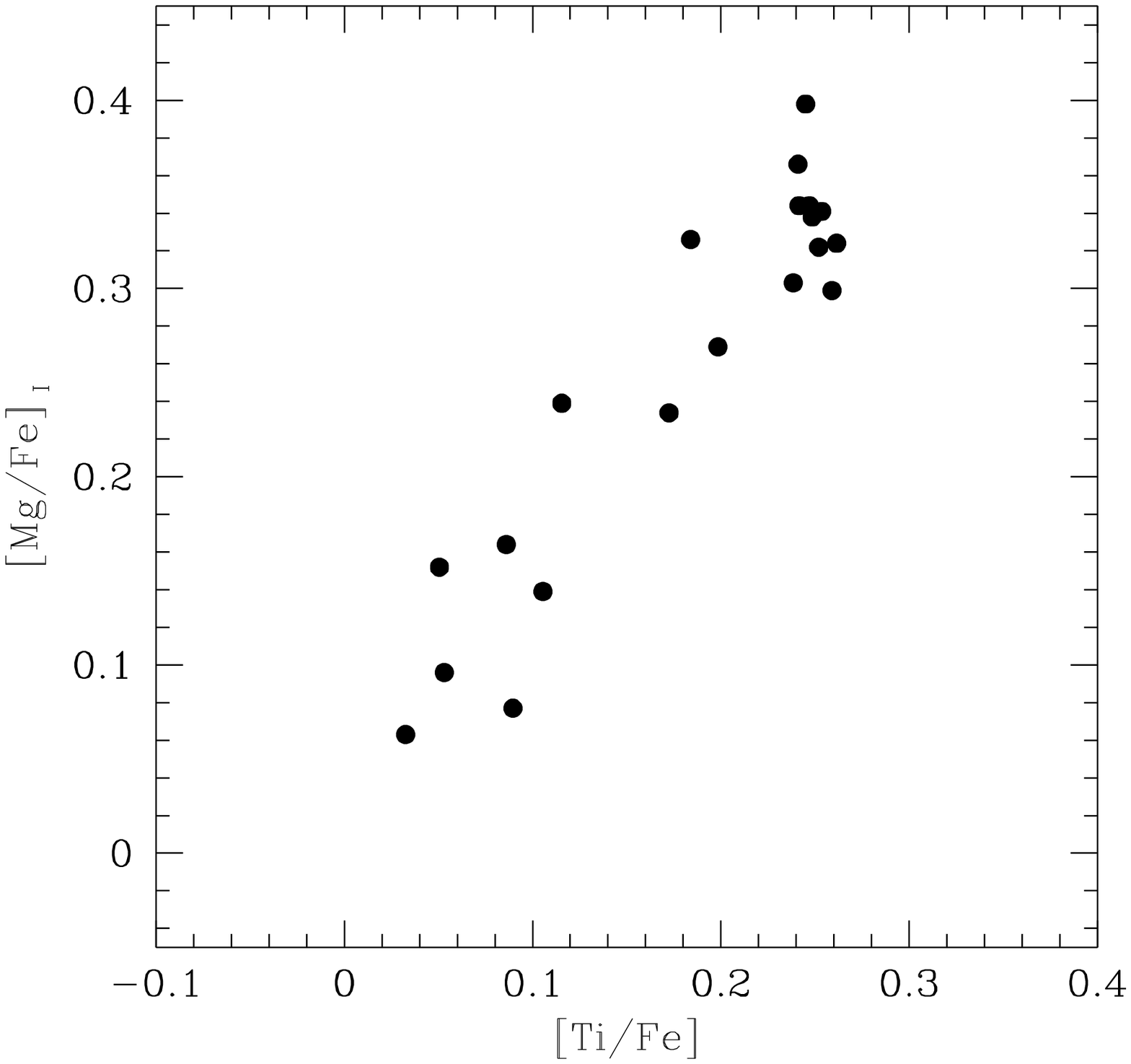}
\end{center}
\end{minipage}
\hfill
\begin{minipage}{8.5cm}
\begin{center}
\leavevmode 
\epsfxsize=8.5cm
\epsfysize=8.5cm 
\epsffile{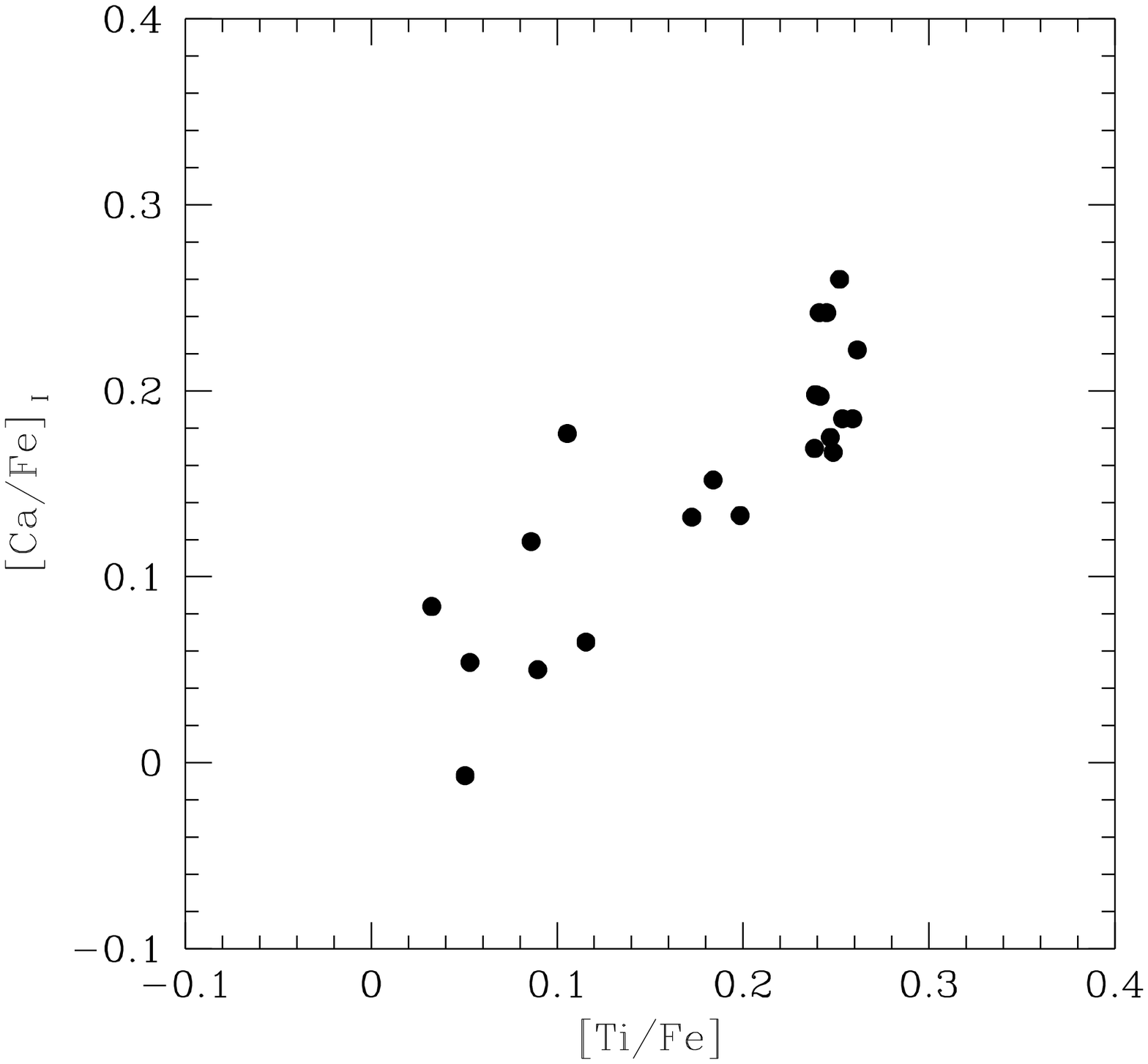}
\end{center}
\end{minipage}
\vspace*{-5mm}
\caption{Correlation diagrams for [Mg/Fe] and [Ca/Fe] versus [Ti/Fe]}
\end{figure*}

\begin{table}
\caption[]{Influence of errors in the model parameters on the HD\,199289 
abundances}
\begin{flushleft}
\begin{tabular}{|lccccc|}
\hline
 Element       &$\Delta T_{\rm eff}$  &$\Delta \log g$  &$\Delta$ [Fe/H]  &$\Delta \xi$  &Rms   \\  
 abundance     &+20K           & +0.2            &+0.2             &+0.1          &sum   \\
\hline
\hline                                
{[Fe/H]}\,I    &+0.014         &+0.010         &$-$0.001         &$-$0.009      &0.019  \\
{[Fe/H]}\,II   &+0.002         &+0.090         &+0.014           &$-$0.008      &0.091  \\
{[Cr/Fe]}\,I   &+0.002         &+0.002         &+0.003           &+0.002        &0.005  \\
{[Cr/Fe]}\,II  &$-$0.003       &+0.002         &$-$0.005         &$-$0.004      &0.007  \\
{[Ti/Fe]}\,I   &+0.002         &$-$0.003       &+0.002           &+0.006        &0.007  \\
{[Ti/Fe]}\,II  &+0.000         &$-$0.003       &$-$0.005         &$-$0.010      &0.012  \\
{[Mg/Fe]}\,I   &+0.010         &+0.020         &$-$0.007         &$-$0.012      &0.026  \\
{[Ca/Fe]}\,I   &$-$0.003       &$-$0.007       &+0.004           &+0.005        &0.010  \\
{[V/Fe]}\,I    &+0.007         &$-$0.005       &+0.002           &+0.009        &0.013  \\
{[Ni/Fe]}\,I   &$-$0.001       &+0.004         &+0.001           &+0.002        &0.005  \\
{[Sr/Fe]}\,I   &+0.003         &$-$0.003       &+0.000           &+0.005        &0.007  \\
{[Y/Fe]}\,II   &+0.006         &+0.004         &+0.005           &$-$0.002      &0.009  \\
{[Zr/Fe]}\,II  &+0.007         &$-$0.012       &+0.008           &+0.007        &0.017  \\
{[Ba/Fe]}\,II  &+0.001         &$-$0.009       &+0.000           &+0.003        &0.010  \\
{[La/Fe]}\,II  &+0.008         &+0.000         &+0.009           &+0.003        &0.012  \\
{[Ce/Fe]}\,II  &+0.009         &$-$0.008       &+0.011           &+0.006        &0.017  \\
{[Nd/Fe]}\,II  &+0.008         &$-$0.013       &+0.017           &+0.004        &0.023  \\
{[Sm/Fe]}\,II  &+0.009         &$-$0.012       &+0.011           &+0.007        &0.020  \\
{[Eu/Fe]}\,II  &+0.009         &$-$0.025       &+0.011           &+0.003        &0.029  \\
\hline
\end{tabular}
\end{flushleft}
\vspace*{-0.75cm}
\end{table}

The variations of the abundance ratios due to changes  on effective 
temperature (+20K), surface gravity (+0.2\,dex), model metallicity 
(+0.2\,dex) and microturbulence velocity (+0.1\,dex) are summarized in 
Table 8 for a typical star.   
The total uncertainty never exceeds some 10\% for all abundance ratios. 

In Fig.~2, we show
the abundance of titanium relative to iron as deduced from the neutral lines
versus [Fe/H]. 
It appears that the scatter in [Ti/Fe],
which amounts to 0.08 dex, is much larger than the estimated error on that 
abundance ratio (0.013).
It seems therefore that the observed scatter is not due to analysis 
uncertainties and could be genuine cosmic scatter.

We confirm this result by comparing the values of [Ti/Fe] deduced from neutral
lines with those obtained from the ionized lines, as shown in Fig.~3. 
We see that the correlation between the two ratios is close to perfect, the 
remaining scatter amounting to 0.025 dex only, which is compatible with the 
estimated errors (0.024 dex). These two determinations of [Ti/Fe] are 
completely independent since the sensitivity of the ionic lines to errors on 
the model parameters is different from that of the neutral lines.

We conclude that the observed scatter is indeed of cosmic origin and that the 
high precision of our results allows us to investigate the correlations 
between different relative abundances.

\section{Abundance correlations}

\subsection{$\alpha$ elements}

In Fig.\  4 we show the abundance of Mg relative to Fe, [Mg/Fe], as a 
function of [Ti/Fe], and [Ca/Fe] versus [Ti/Fe]. These two figures show that 
the three $\alpha$-elements Mg, Ca and Ti behave in the same way, which 
means that they were synthesized by the same nucleosynthetic process in 
similar objects. This is in agreement  with the generally accepted view that 
the $\alpha$-elements are mainly produced in supernova explosions of massive 
stars (Arnett 1991, Thielemann et al.\  1993, Woosley and Weaver 1995).

\subsection{Iron peak elements}

\begin{figure*}
\begin{minipage}{8.5cm}
\begin{center}
\leavevmode 
\epsfxsize=8.5cm
\epsfysize=8.5cm 
\epsffile{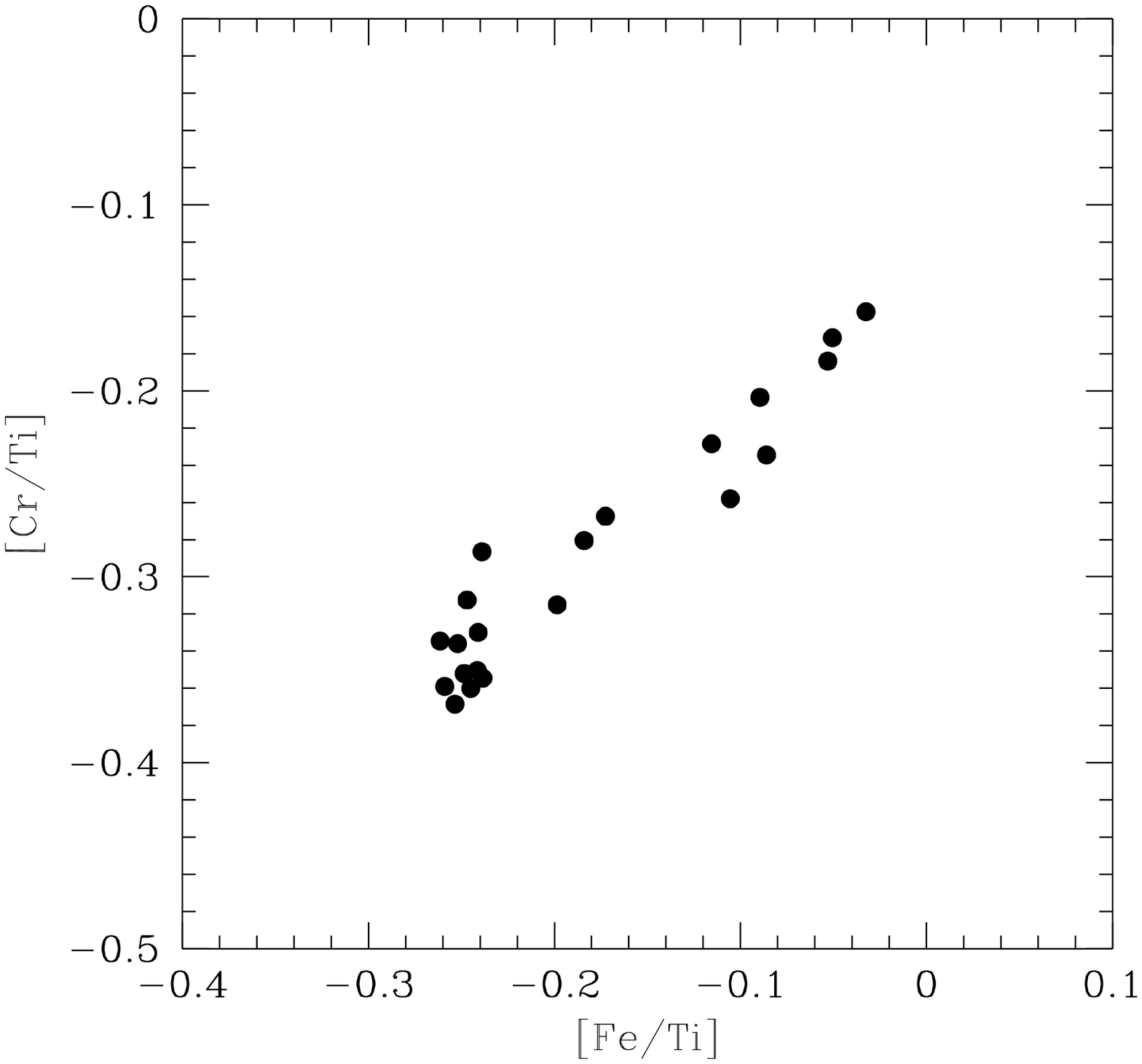}
\end{center}
\end{minipage}
\hfill
\begin{minipage}{8.5cm}
\begin{center}
\leavevmode 
\epsfxsize=8.5cm
\epsfysize=8.5cm 
\epsffile{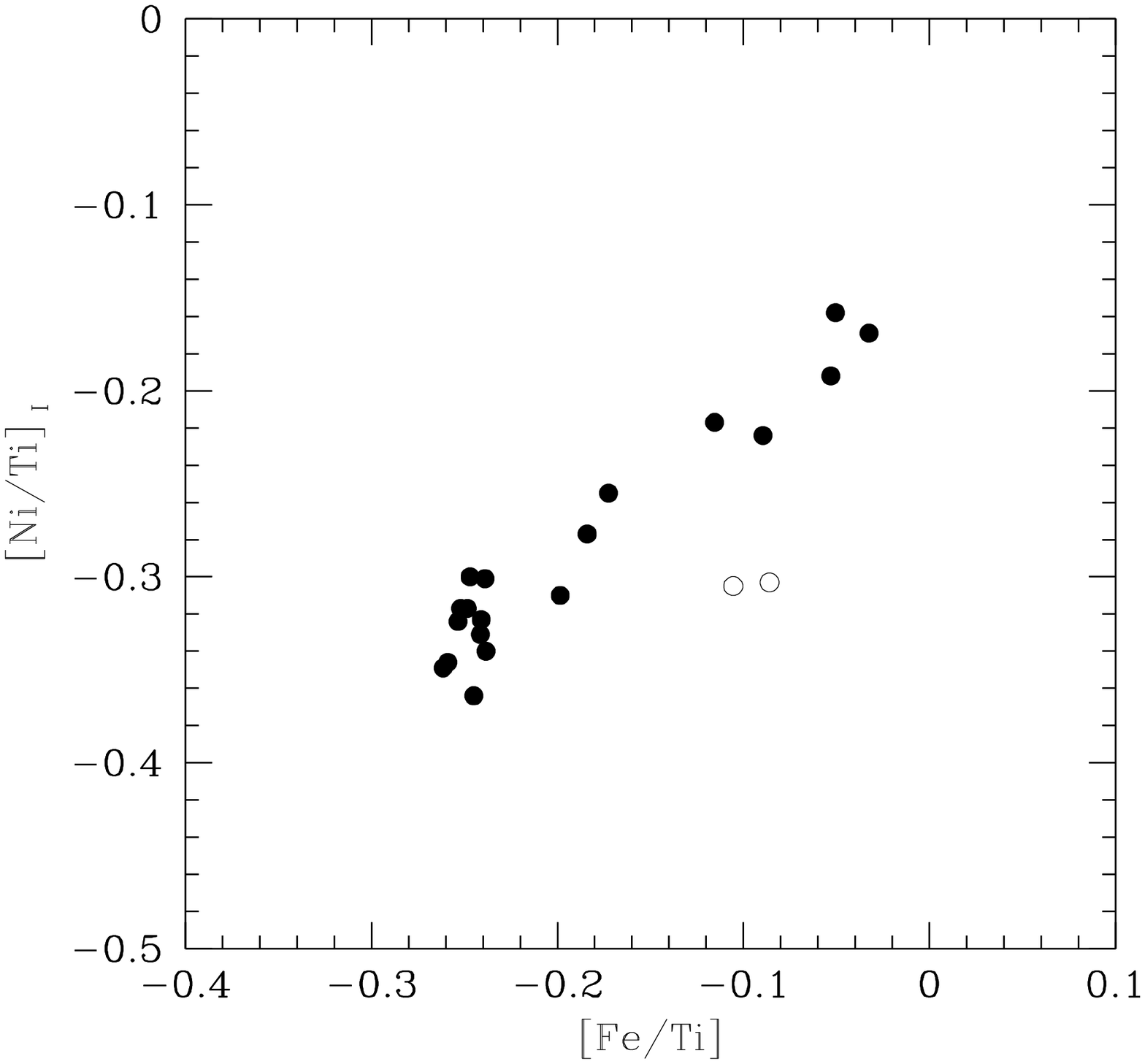}
\end{center}
\end{minipage}
\vspace*{-5mm}
\caption{Correlation diagrams for [Cr/Ti] and [Ni/Ti] versus [Fe/Ti]}
\end{figure*}

The abundances of Cr and Fe relative to Ti are displayed in Fig.\  5. The 
correlation is very good, with a scatter of only 0.023 dex. The second part 
of the same figure shows the same comparison for Ni and Fe. Here again, the 
correlation is remarkable ($\sigma$ = 0.013) for all but two stars, which 
appear to be somewhat depleted in nickel. These two stars, which also display 
other abundance anomalies (as we shall see below) are identified by open 
symbols in this figure and the following ones.

While the even iron peak elements correlate very well with each other, the 
situation appears different for the odd element vanadium. As shown in 
Fig.\  6, the vanadium abundance does not correlate well with the other iron 
peak ones.  The correlation is only slightly better with  the 
$\alpha$-elements. It is difficult to draw any conclusion without  
investigating the behaviour of the other odd iron peak elements Mn and Co.

\begin{figure*}
\begin{minipage}{8.5cm}
\begin{center}
\leavevmode 
\epsfxsize=8.5cm
\epsfysize=8.5cm 
\epsffile{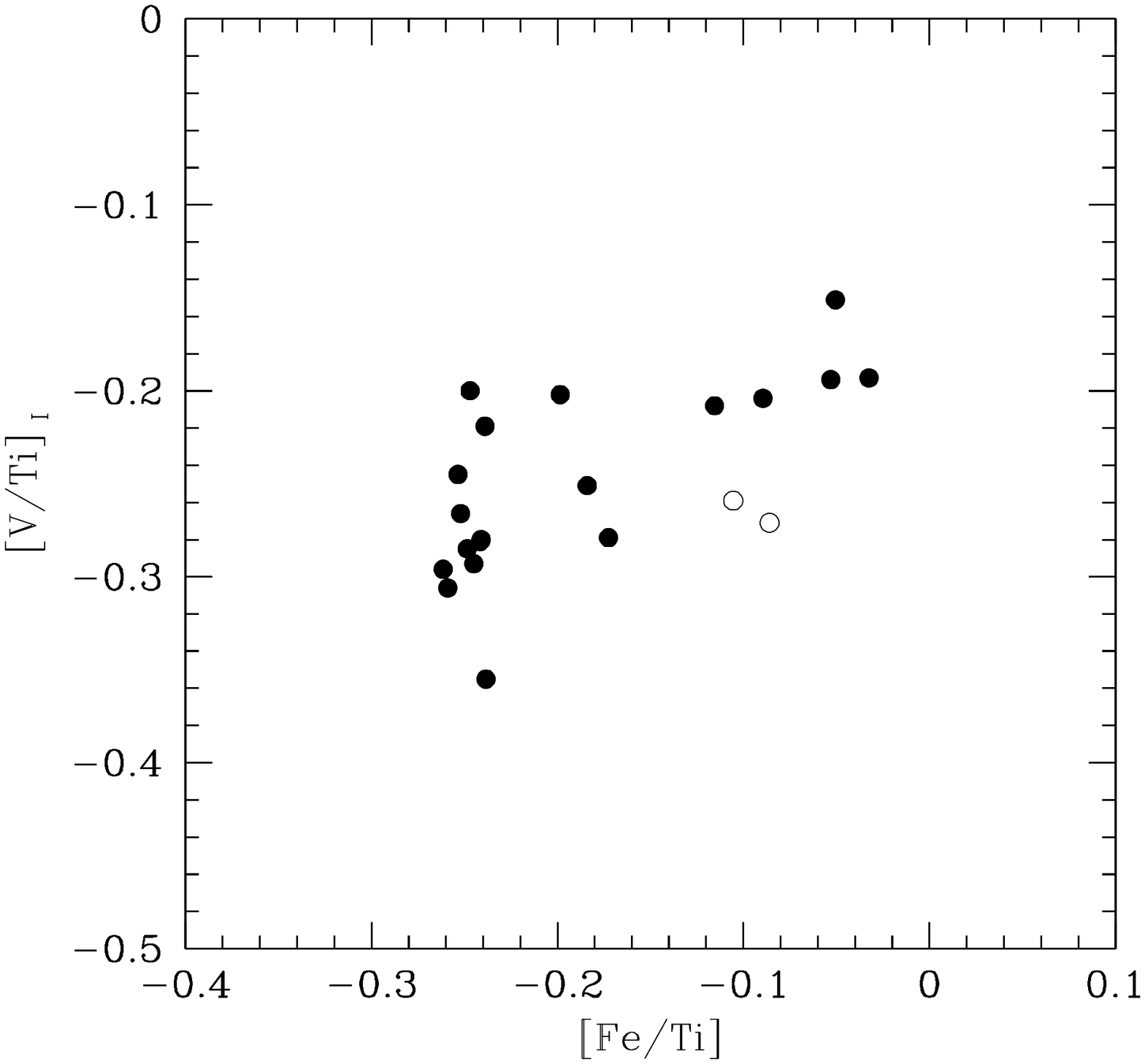}
\end{center}
\end{minipage}
\hfill
\begin{minipage}{8.5cm}
\begin{center}
\leavevmode 
\epsfxsize=8.5cm
\epsfysize=8.5cm 
\epsffile{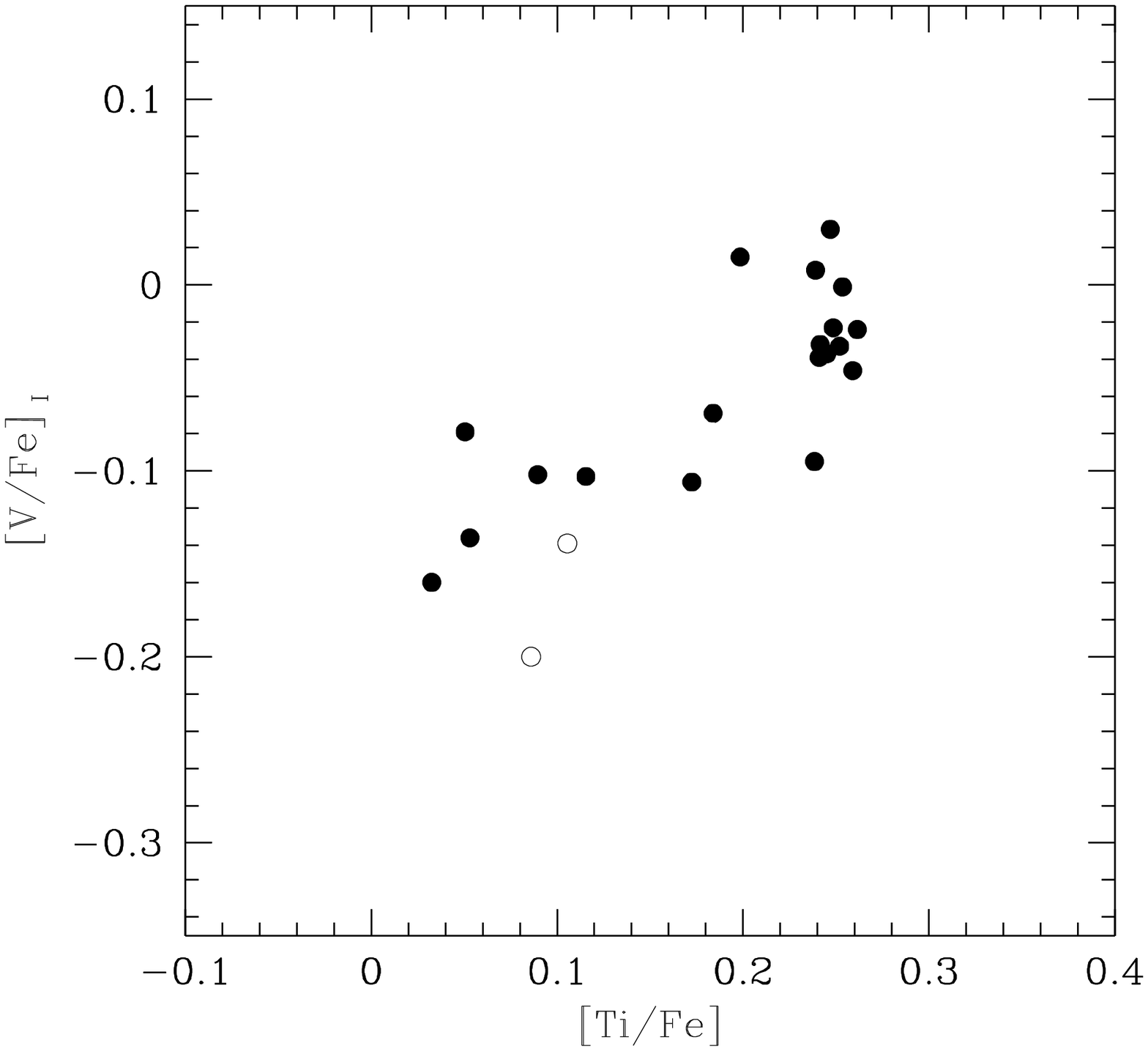}
\end{center}
\end{minipage}
\vspace*{-5mm}
\caption{Correlation diagrams for [V/Ti] versus [Fe/Ti]
              and [V/Fe] versus [Ti/Fe]} 
\end{figure*}

\subsection{Neutron capture elements} 

We wanted to carry out the same analysis for the neutron capture elements, 
in order to distinguish between the $s$ and $r$ processes at an early stage 
of the galactic 
evolution, and to identify the most likely sites for their formation.
A correlation plot for a typical $s$ element, yttrium, is shown in 
Fig.\  7, together with the corresponding diagram for the $r$ element 
europium. The difference between the behaviours of Y and Eu is striking. 
Apart for the two ``anomalous" stars, there is a one-to-one correlation 
between the $r$-process element Eu and the $\alpha$-element Ti. All points 
are located on a single straight line with a slope close to 1 
and ending with a 
clumping at the maximum value of [Ti/Fe].

\begin{figure*}
\begin{minipage}{8.5cm}
\begin{center}
\leavevmode 
\epsfxsize=8.5cm
\epsfysize=8.5cm 
\epsffile{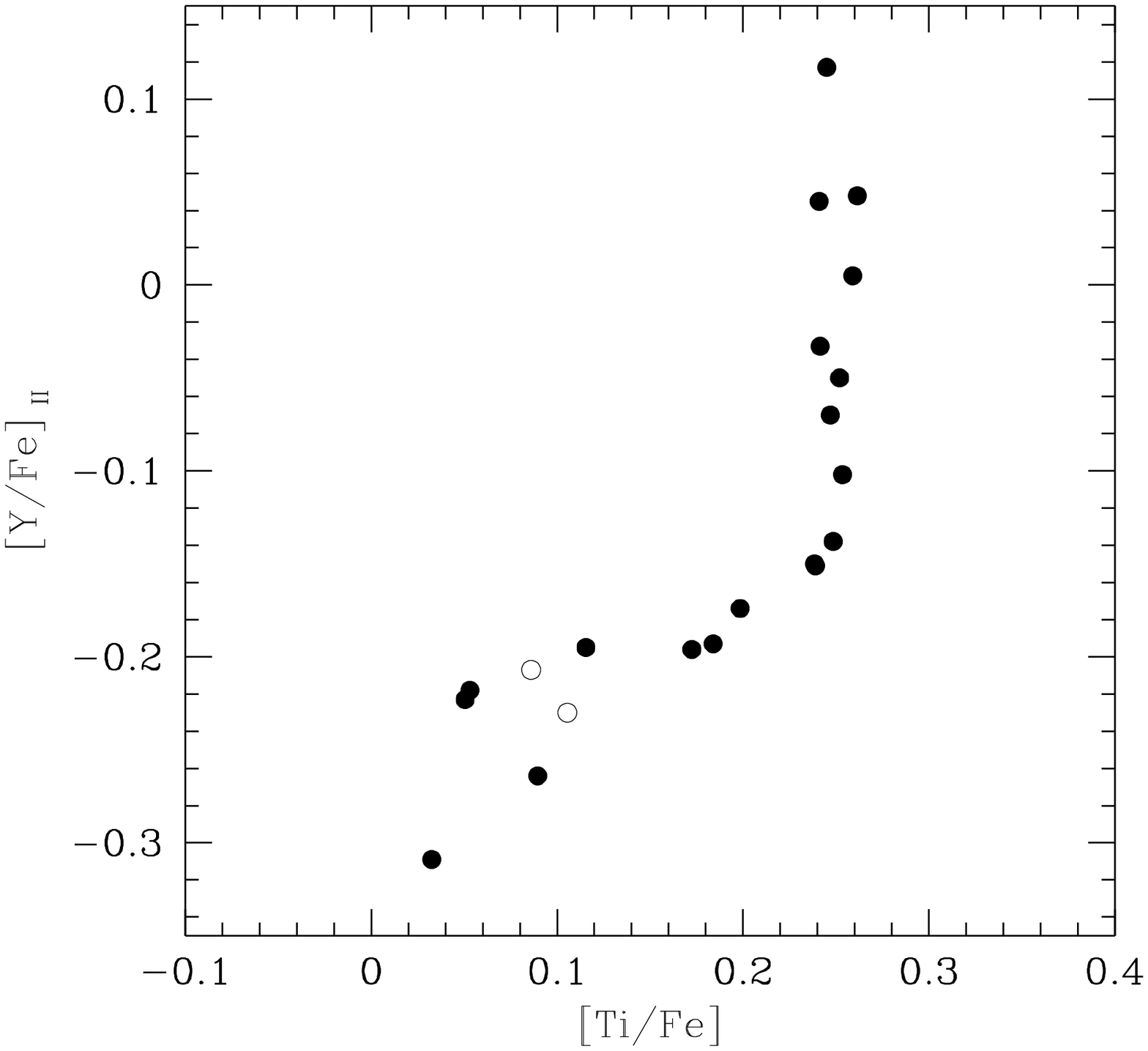}
\end{center}
\end{minipage}
\hfill
\begin{minipage}{8.5cm}
\begin{center}
\leavevmode 
\epsfxsize=8.5cm
\epsfysize=8.5cm 
\epsffile{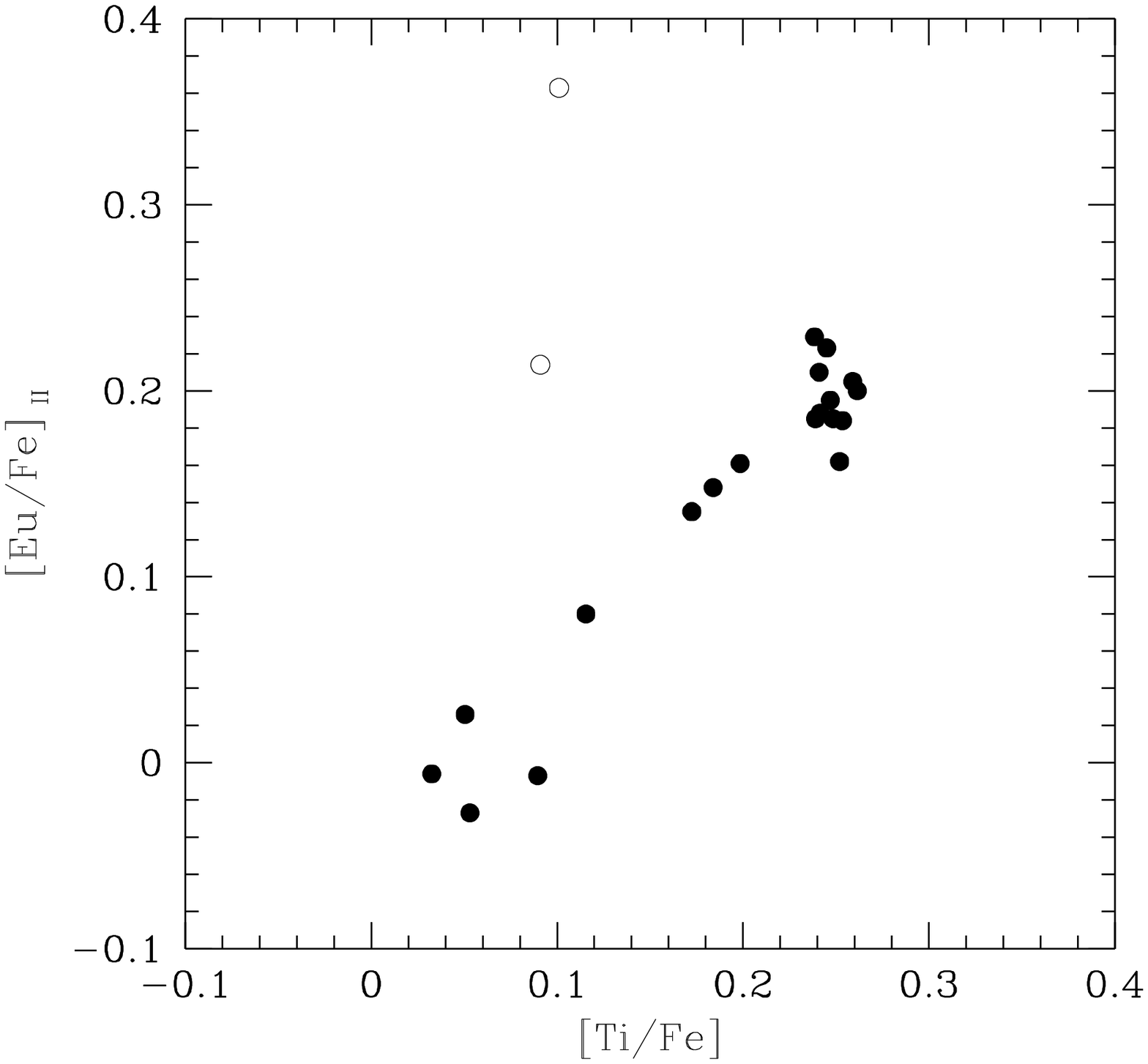}
\end{center}
\end{minipage}
\vspace*{-5mm}
\caption{Correlation diagrams for [Y/Fe] and [Eu/Fe] versus [Ti/Fe]}
\end{figure*}

\begin{figure*}
\begin{minipage}{8.5cm}
\begin{center}
\leavevmode 
\epsfxsize=8.5cm
\epsfysize=8.5cm 
\epsffile{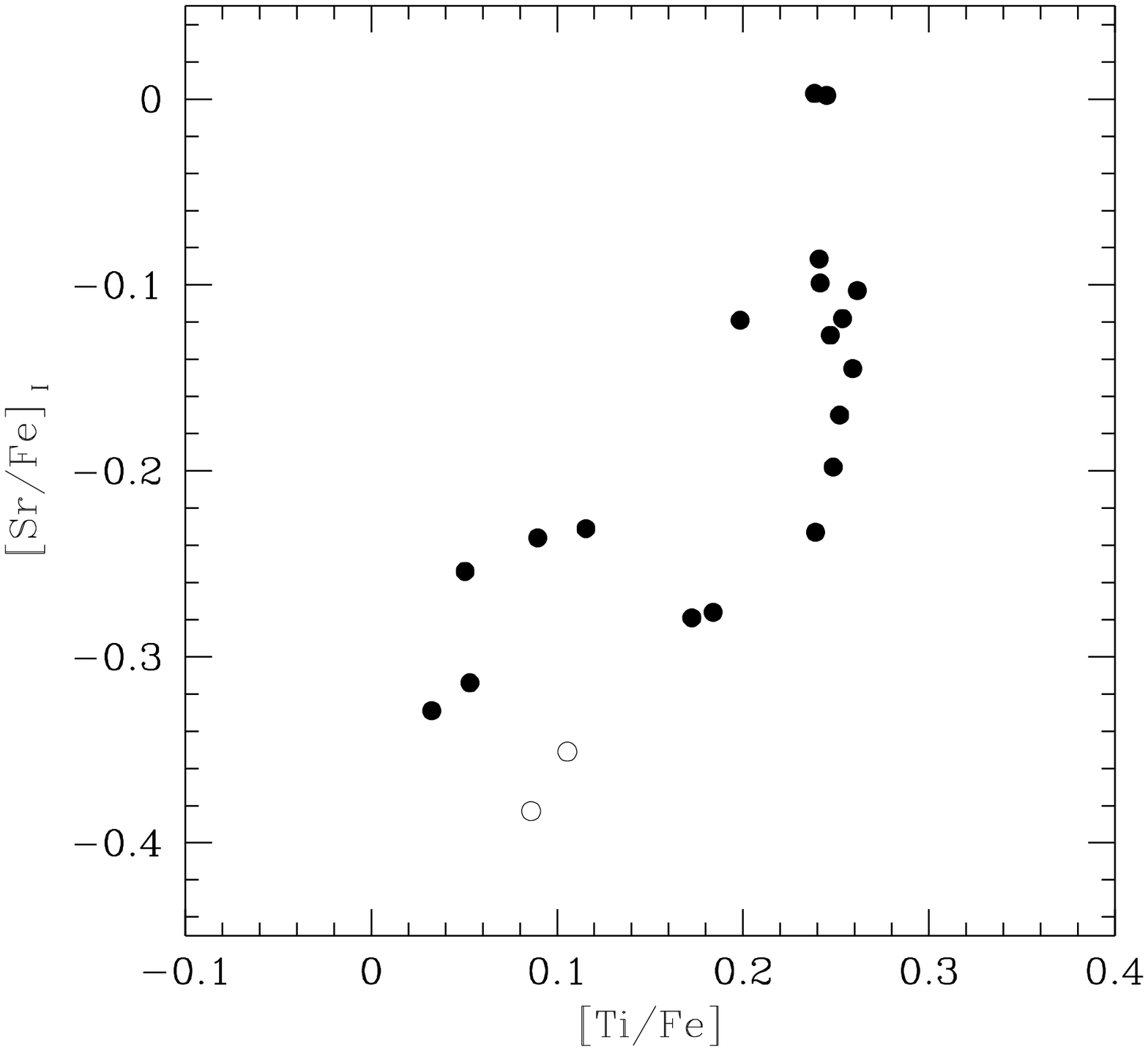}
\end{center}
\end{minipage}
\hfill
\begin{minipage}{8.5cm}
\begin{center}
\leavevmode 
\epsfxsize=8.5cm
\epsfysize=8.5cm 
\epsffile{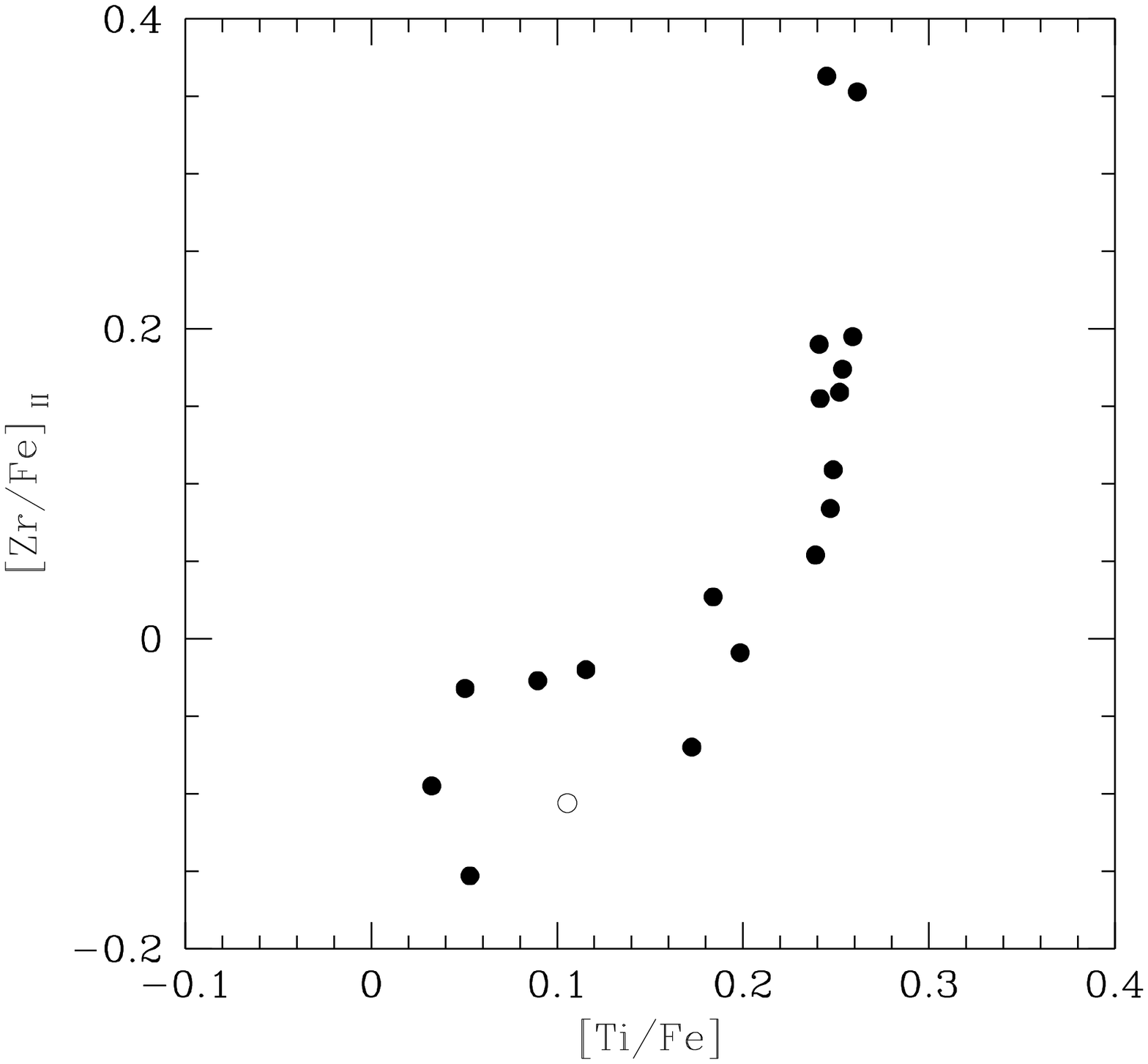}
\end{center}
\end{minipage}
\vspace*{-5mm}
\caption{Correlation diagrams for [Sr/Fe] and [Zr/Fe] versus [Ti/Fe]}
\end{figure*}

In contrast, the diagram of the $s$-process element Y is much more complex. 
Here, we do not have a simple correlation with the $\alpha$-elements, but 
rather, two 
distinct behaviours. For about half of the stars, corresponding to low values 
of [Ti/Fe], there is a correlation between [Y/Fe] and [Ti/Fe], but the 
slope is significantly smaller than 1.

The remaining stars have a constant (and maximum) [Ti/Fe] and increasing 
values of 
[Y/Fe], starting at the maximum value reached by the first group. We refer 
to such an abundance correlation as a {\it two branches diagram}.

Such a strikingly different behaviour needs to be confirmed by other 
$r$ and $s$ elements. Figure 8 shows the same correlation diagram for the 
other light $s$-process elements Sr and Zr, which agree very well with the 
diagram for Y, although the Sr and Zr results are based on the analysis of 
a single line, which is moreover very weak in the case of Zr. The three 
light $s$-process elements Sr, Y and Zr thus show a common pattern. 
We propose an explanation for these {\it two branches diagrams} in Section 9.

\begin{figure}
\begin{center}
\leavevmode 
\epsfxsize= 8.5 cm 
\epsffile{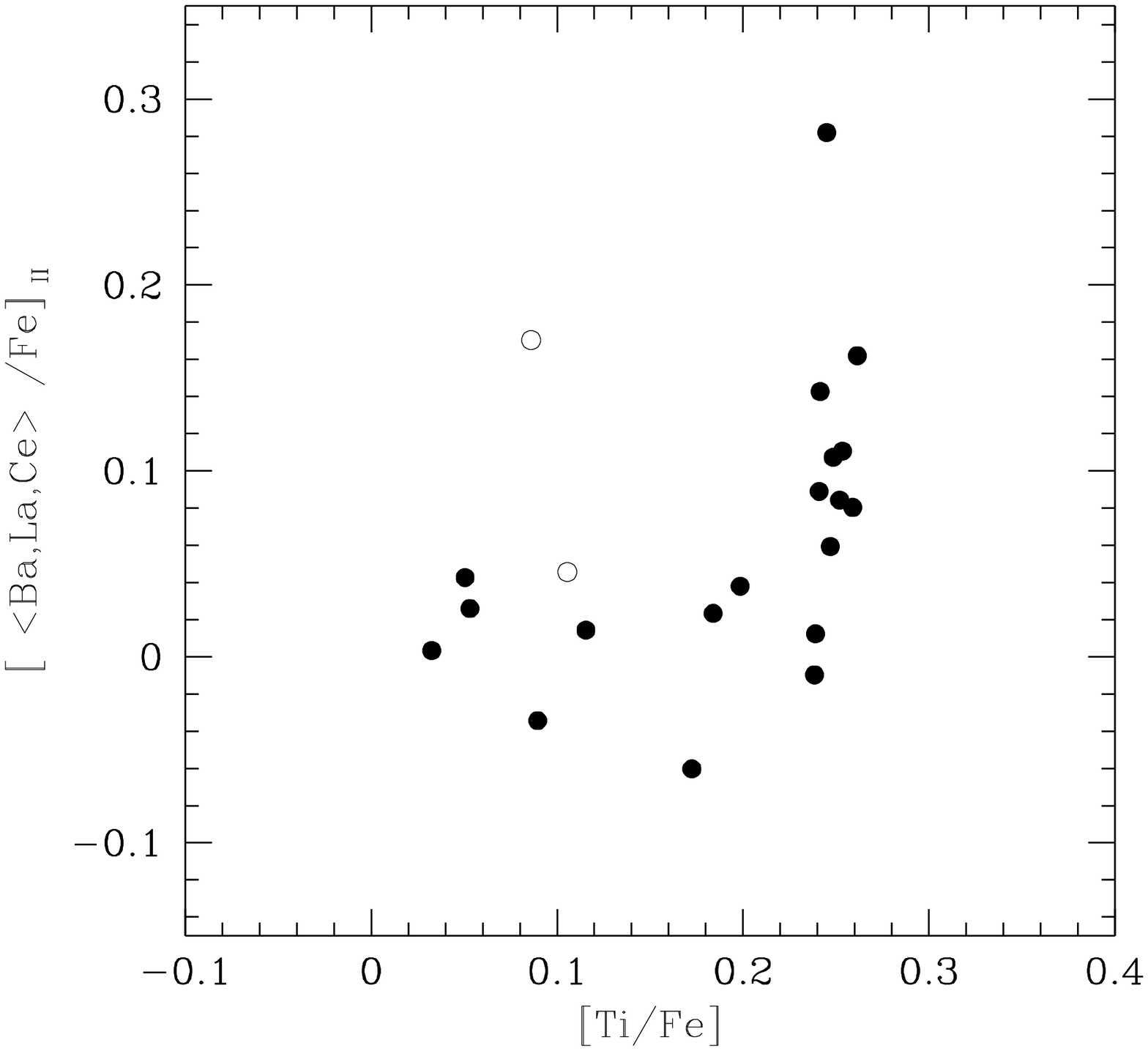}
\vspace*{-5mm}
\caption{Correlation diagram for [$<Ba,La,Ce>$/Fe] versus [Ti/Fe]}
\end{center}
\end{figure}

\begin{table*}
\caption[]{Kinematic data.  Radial velocity $v_R$, parallax $\pi$, proper 
motion in right ascension $\mu_\alpha$ and in declination
$\mu_\delta$, and the deduced galactic velocity components with respect to 
the LSR: U, V, W. (U is positive towards the galactic anticentre)}
\begin{tabular}{|crrrrrrr|}
\hline
   ID	    & \multicolumn{1}{c}{$v_R$}    &\multicolumn{1}{c}{$\pi$}  &\multicolumn{1}{c}{$\mu_\alpha$}       &\multicolumn{1}{c}{$\mu_\delta$}      &\multicolumn{1}{c}{U}      &\multicolumn{1}{c}{V}       &\multicolumn{1}{c}{W}   \\
	    & (km/s)         &   (mas)     & (mas/yr)  & (mas/yr)   & (km/s)  & (km/s) & (km/s)\\     
\hline
\hline
1     & 118.0         & 41.07    & 689.67    &$-$214.34   &  98.36   & $-$70.97  &$-$37.11   \\
2     &  52.0         & 19.02    & 347.27    &  413.51    & 118.63   & $-$49.50  &   1.44   \\
3     &  57.4         & 33.40    &$-$92.20   &$-$167.52   & 19.33   & $-$36.24  &$-$10.81   \\
4     &$-$1.0         & 12.01    &$-$12.86   &$-$201.75   &$-$80.07  &   17.71   &$-$31.50   \\
5     & 102.6         & 65.79    &$-$220.83  & 1722.89    & 135.33   & $-$43.40  &  46.89   \\
6     &  62.0         & 20.14    & 152.49    &$-$239.09   &$-$39.46  & $-$71.50  &   9.89   \\
7     & 120.8         & 46.90    & 244.35    &  213.46    &  38.23   & $-$76.80  &  77.39   \\
8     &$-$1.0         & 25.16    &$-$64.62   &$-$132.59   &$-$20.41  &   16.88   &$-$18.56   \\
9     &  32.0         & 18.66    &$-$157.82  &$-$56.85    &   9.35   & $-$18.72  &$-$29.12   \\
10    &  51.1         & 17.77    & 159.19    &$-$201.28   &$-$83.37  & $-$5.22  &$-$30.28   \\
11    &$-$10.0        & 15.34    &$-$45.01   &   79.26    &   7.83   &   17.55   &  30.87   \\
12    &$-$9.0         & 17.71    &$-$12.45   &$-$64.29    &   2.99   &   15.15   &$-$7.15   \\
13    &  28.5         & 16.80    &   0.71    &$-$14.74    &$-$26.66  &   12.72   &  22.12   \\
14    &  31.0         & 13.72    &$-$63.25   &  17.68     &$-$34.98  & $-$11.53  &  20.66   \\
15    &$-$14.7        & 14.76    &$-$309.16  &$-$365.29   &$-$84.12  &$-$113.76  &  50.62   \\
16    &$-$172.0       & 22.88    & 539.73    &$-$1055.93  & 145.69   &$-$230.16  &$-$66.03   \\
17    &$-$247.0       & 17.94    & 117.90    &$-$549.71   &  66.11   &$-$262.10  &$-$23.53   \\
18    &$-$30.0        & 15.78    &  44.38    &$-$428.55   &$-$12.17  &$-$113.75  &$-$26.24   \\
19    &$-$16.5        & 18.94    & 169.40    &$-$283.37   &  37.84   & $-$51.77  &$-$11.21   \\
20    &$-$29.4        &108.50    &  81.08    &  800.68    &   2.69   &   58.83   &  13.47   \\
21    &$-$33.6        & 23.66    & 150.64    &  331.61    &  55.35   &   29.86   &  52.98   \\
\hline
\end{tabular}
\end{table*}

The heavy $s$-process elements Ba, La and Ce have been considered together 
in Fig.\  9 since the available spectroscopic data were of slightly lower 
quality.
Despite the larger scatter,  these elements all follow the same  trend. 
Once again, 
we can distinguish two groups of stars, the first one with variable [Ti/Fe] 
and a second vertical one at the maximum value of this abundance ratio.

The slope for the first group of stars is not as well defined as in the case 
of the lighter $s$ elements. This may be due to the lower quality of the 
data, but might also 
reflect a different nucleosynthetic history. The latter hypothesis is 
supported by the fact that at least one of the anomalous stars, which 
perfectly fits in the light $s$ diagrams, now shows up again as overenriched 
in heavier $s$ elements.

The cases of Nd and Sm are illustrated in Fig.\  10. These elements, 
formed by both neutron capture processes in the solar system, display a 
pattern intermediate between those of the $r$-process element Eu and the 
heavy $s$-process ones. For low values of [Ti/Fe], [Nd/Fe] and [Sm/Fe] are 
roughly constant, except for the two anomalous stars, which are relatively 
enriched. Near the maximum value of [Ti/Fe], a looser clumping appears, with 
a dispersion which is 
intermediate between those of the $r$ and $s$ elements.

As we have seen earlier, the two stars HD\,193901 and HD\,194598 
represented by 
open symbols in the correlation diagrams, display a number of abundance 
anomalies: a lower Ni abundance (Fig.\  5b), a probable overabundance in
heavy $s$-process elements (Fig.\  9) and, above all, a higher abundance
in $r$-process products (Figs.\  7b, 10).  If, as stated by Anders and
Grevesse (1989), Ni is the even iron peak element with the smallest 
contribution
from explosive nucleosynthesis, this behaviour might be explained by a
different nucleosynthetic history, namely an enhanced contribution from 
explosive processes compared to equilibrium ones.

\begin{figure*}
\begin{minipage}{8.5cm}
\begin{center}
\leavevmode 
\epsfxsize=8.5cm
\epsfysize=8.5cm 
\epsffile{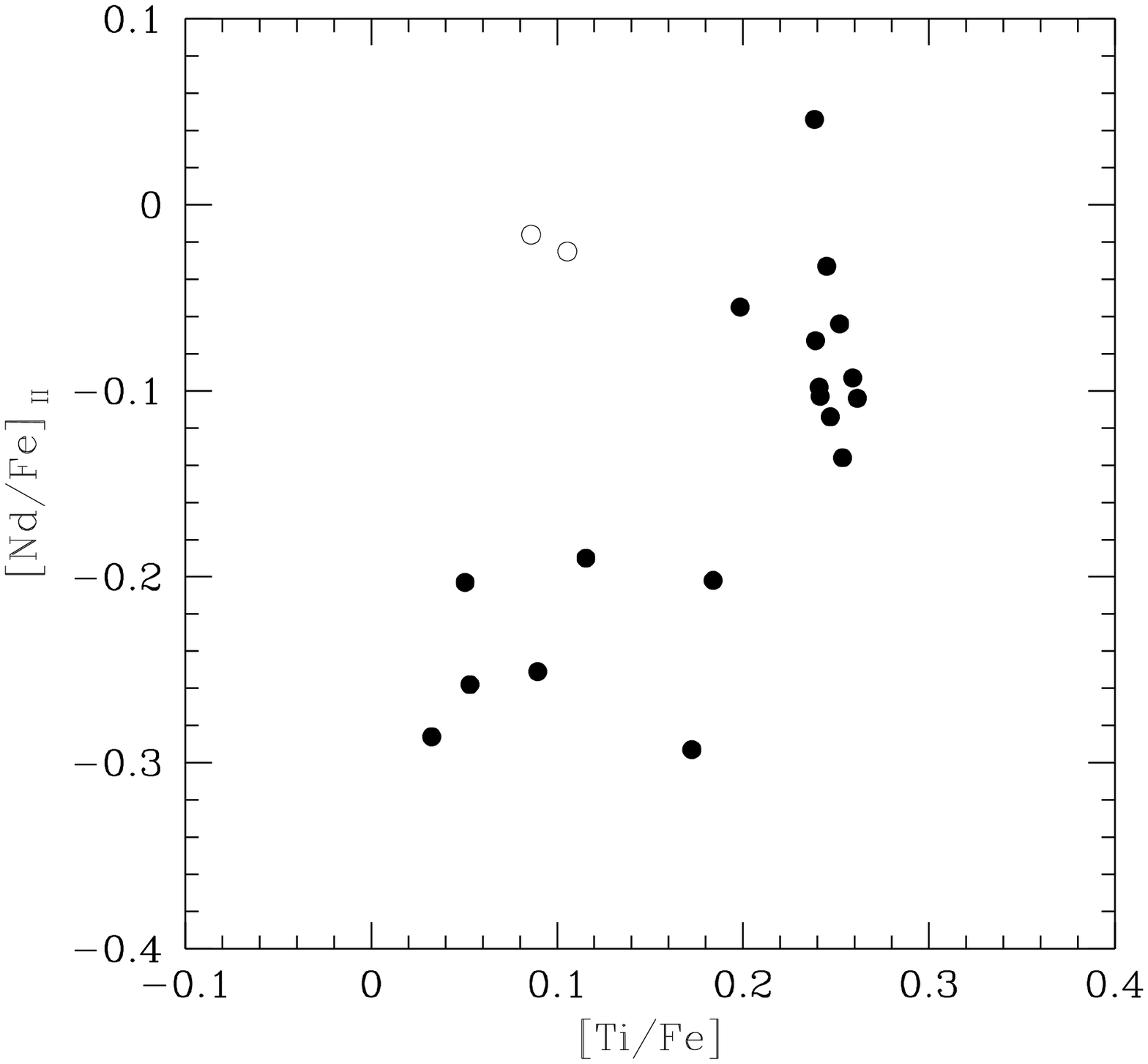}
\end{center}
\end{minipage}
\hfill
\begin{minipage}{8.5cm}
\begin{center}
\leavevmode 
\epsfxsize=8.5cm
\epsfysize=8.5cm 
\epsffile{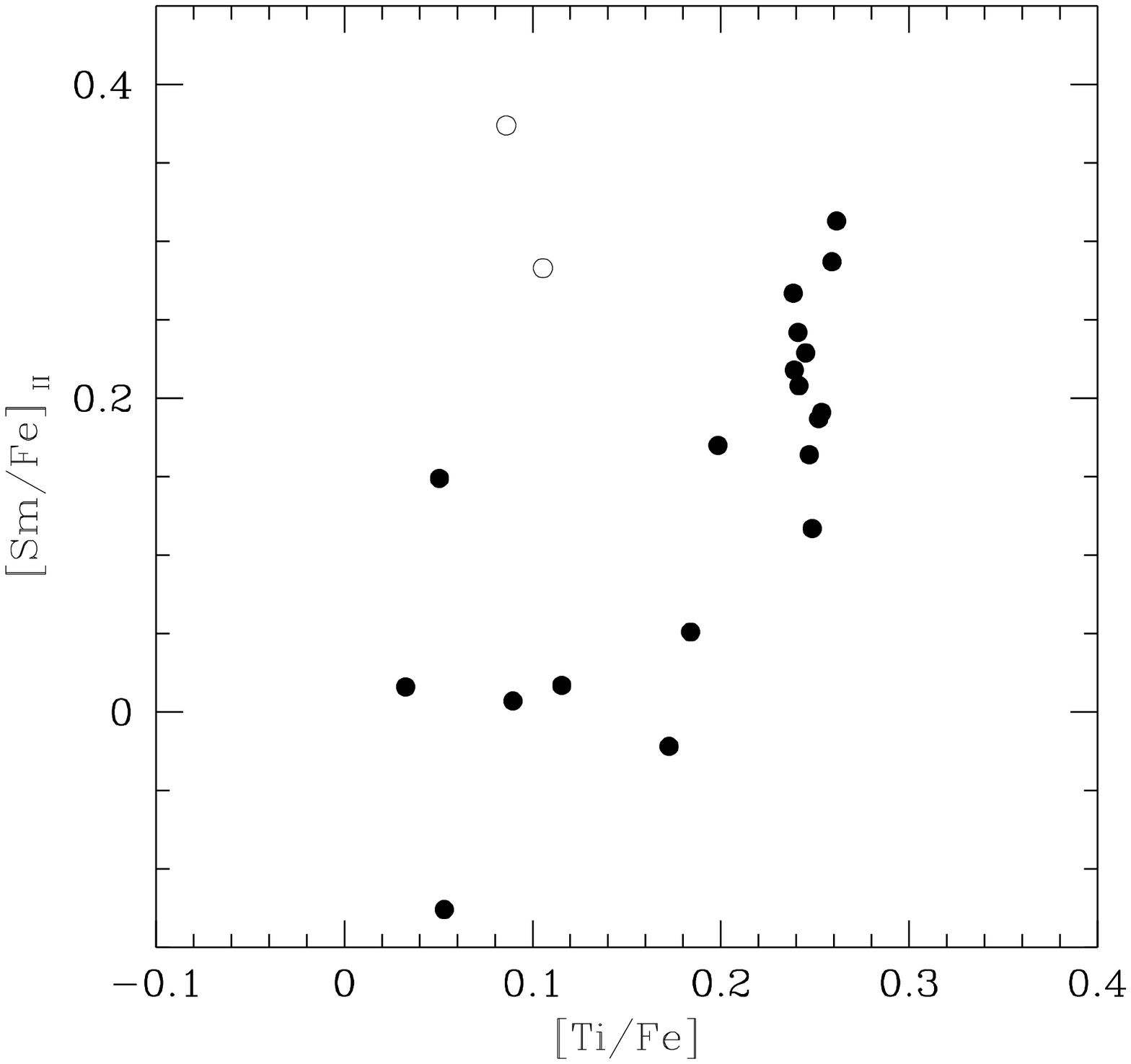}
\end{center}
\end{minipage}
\vspace*{-5mm}
\caption{Correlation diagrams for [Nd/Fe]  and [Sm/Fe] versus [Ti/Fe]}
\end{figure*}

\section{Kinematics of the stars}

The kinematical data are summarized in Table 9.
The radial velocities were selected from several sources:  
Barbier-Brossat et al.\   (\cite{BB94}),  Wilson (\cite{GCRV}), 
Evans (\cite{E}),  Gratton and Sneden (\cite{GS91}), Lindgren and Andersen 
(private communications). 
The radial velocities are known to about $\pm\,2\,$km\,s$^{-1}$ for most 
of the stars.
The proper motions and parallaxes used to calculate the galactic 
velocities were taken from the Hipparcos Catalogue (\cite{ESA}).
The calculation of the galactic space velocities U, V and W with respect 
to the LSR is based on the method presented in Johnson and Soderblom 
(\cite {JS87}). The corrections applied to the observed velocities for the 
solar motion are 
\newline
(-10.4,+14.8,+7.3) km\,s$^{-1}$ in (U,V,W) 
(Mihalas and Routly 1968).

Upon examination of Table 9, our sample appears to contain thick disk and 
halo stars. However, there is no clear distinction between these two 
populations in Figs.\ 7a, 8 and 9, since thick disk stars are found on both 
branches in these plots.

This result seems to disagree with the analysis of Nissen and Schuster 
(1997, NS), who have selected two samples of stars on the basis of their 
kinematical properties and covering about the same metallicity range as ours. 
While their halo stars are indeed found in both branches of our diagram, all 
their disk stars display maximum [$\alpha$/Fe] values. 
Some of our disk stars undoubtedly have low [$\alpha$/Fe].
The apparent disagreement might be due to the fact that both samples contain 
a relatively small number of stars and that, by accident, all disk stars 
selected by NS have a high
[$\alpha$/Fe].

It may be of interest to note that the two stars which seem to display some
anomalous abundances are just  these which have the most halo-like 
kinematic properties.  However, it is difficult to conclude on the basis
of only two stars.  We plan to extend the analysis to other metal-poor
stars with kinematics more typical of the halo.

\section{The ``Two branches diagram"}

\subsection{Universality of the ``Two branches diagram"}

The analysis of the correlations between the relative abundances of a
number of elements for stars with roughly 0.1 of the solar metallicity
leads us to distinguish between two stellar
populations, corresponding to distinct branches in the diagram 
(e.g.\  Fig.\  7a): 
(a) a fraction of the stars have a range
of moderate overabundances of the $\alpha$-elements and either a
constant or slowly varying abundance of the $s$-process elements
relative to the iron peak, and 
(b) the others show a constant (and maximum) overabundance
of the $\alpha$-elements relative to the iron-peak elements,
and a range in $s$-process abundances.
This behaviour must be related to nucleosynthesis processes.

Since our sample contains a limited number of stars, its astrophysical
significance may be checked by including the results of other analyses.
In Fig.\  11, we have added the NS stars as well as the sample of Zhao and 
Magain (1991, ZM). The quality of those data is almost as good as the quality 
of the data presented here.  The ZM data extend our sample towards stars of 
lower metallicity.  The stars in Fig.\ 11 have metallicities
$-2 \leq$ [Fe/H] $\leq -0.6$. 
The zero points have been fixed by comparing
the results obtained for the stars in common (3 with NS, 10 with ZM), for 
which we have kept the results obtained here with a more precise analysis.  
We see that  all these metal-poor stars follow the same trend, independently 
of their metallicity.

\begin{figure}
\begin{center}
\leavevmode 
\epsfxsize= 8.5 cm 
\epsffile{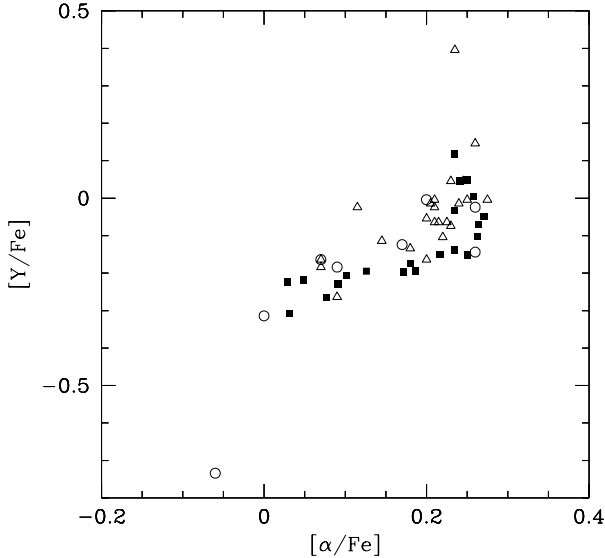}
\vspace*{-5mm}
\caption{Correlation diagram for [Y/Fe] versus [$\alpha$/Fe] with our data
(full squares), the data of Zhao and Magain (1991) (open circles) and the data
of Nissen and Schuster (1997) (open triangles).  
Only stars with [Fe/H] $< -0.6$ are plotted}
\end{center}
\end{figure}

As our metallicity range for (thick) disk stars is rather limited towards 
metal-poor stars, we have added in Fig.\  12 the data obtained by 
Edvardsson et al.\  (1993) for disk stars of various metallicities.  
The 
difference in zero-points between the two analyses has been corrected on the 
basis of 10 stars in common. 
We see that these disk stars definitely {\it do not} follow the relation 
obtained for metal-poor stars, but scatter
mostly through the upper left part of the diagram.  Taking into account the 
larger error bars in Edvardsson et al.\  (1993) data, our metal-poor stars 
define the lower right envelope of this domain.  This envelope also contains 
the lowest metallicity stars of Edvardsson et al.\  (1993).

\begin{figure}
\begin{center}
\leavevmode 
\epsfxsize= 8.5 cm 
\epsffile{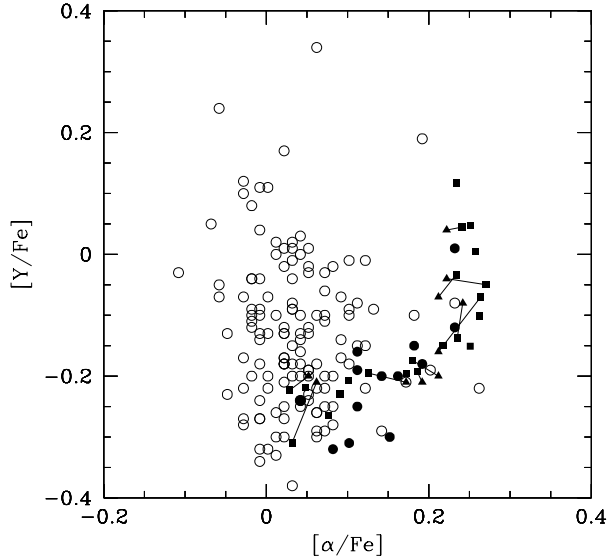}
\vspace*{-5mm}
\caption{Correlation diagram for [Y/Fe] versus [$\alpha$/Fe] with our data 
         (full squares), the data of Edvardsson et al.\   (1993) 
         for [Fe/H] $> -0.6$ (open circles),  for [Fe/H] $< -0.6$ (full
         circles) and  for the stars also included in our work 
         (full triangles), joined by lines to their counterparts}
\end{center}
\end{figure}

On the basis of these two comparisons, a cutoff in metallicity at 
[Fe/H] $\sim -0.6$ seems to emerge.  For stars of higher metallicity, the 
{\it two branches diagram} does not apply.  Lower metallicity stars follow a 
universal relation described by the {\it two branches diagram}.  We shall 
refer to the slowly varying branch stars as Population IIa and to the vertical
branch as Population IIb.

\subsection{Interpretation}

\subsubsection{General picture}

We now briefly describe a possible scenario for the formation and origin 
of these metal-poor stars, which provides an explanation for the universal 
relation described above.
This scenario will be developed in more details in a forthcoming paper.
It takes place in the early stages of the chemical evolution of the
Galaxy. We 
distinguish two separate phases in the evolution of the metal-poor stars.

\vspace{1ex}
\noindent
{\em Phase 1}

At first, we assume that there is a burst of star formation, consisting 
essentially of massive stars.  
As these stars evolve and become supernovae (SNe), $\alpha$-elements and 
$r$-process elements are formed and then ejected in the surrounding 
interstellar matter (ISM).  
New stars are formed from this ISM, which is enriched in those $\alpha$ and 
$r$-process elements. These stars correspond to Pop IIa.

\vspace{1ex}
\noindent
{\em Phase 2}

As time goes on, the lower mass stars are either still contracting towards 
the main sequence or have reached a more evolved phase, maybe already 
processing $s$ elements. These evolved
stars are known to have strong stellar winds
and to undergo superwind events. 
The newly formed $s$ elements are thus ejected into the surrounding ISM, 
previously enriched in $\alpha$- and $r$-process elements.
New born stars formed from this ISM will keep a constant value for
[$\alpha$/Fe] but will be enriched in $s$ elements.  They will belong
to Pop IIb.  

Another possibility is that already formed lower mass stars can accrete gas
from the $s$-process enriched winds.  In this case, the $s$-process enrichment
will show up in the external layers only, at least if the convective envelope
is not too large.  These stars will also belong to Pop IIb, although their
internal chemical composition will be that of Pop IIa at the bottom of the 
vertical branch. 

\subsubsection{Globular cluster scenario}

In this subsection, we will be more specific as to the environment in which
our two-phases scenario could occur.

As we have shown in Section 9.1, the {\it two branches diagram} seems to
define a universal relation for stars more metal-poor than [Fe/H] $\sim -0.6$.
In particular, it should therefore apply to all halo stars.

Since the halo is also populated by globular clusters 
(GCs, see the recent review
by Meylan \& Heggie 1997), we have searched
for a connection between them and the field halo stars (FHS), and we propose
the following EASE (Evaporation/Accretion/Self-Enrichment) scenario:\\
(1) all FHS were born in (proto-) GCs;\\
(2) GCs have undergone a chemical evolution;\\
(3) some of the GCs were disrupted at an early stage in their evolution, the
lower mass stars forming Pop IIa;\\
(4) in the GCs which have survived, accretion of matter from AGB stars 
modifies the surface composition of the cluster stars;\\
(5) some low mass stars evaporate from the GCs or get dispersed in the halo
when the GC is disrupted. They form Pop IIb FHS.

The relation between the thick disk and field halo stars on the one hand and 
GCs on the other hand is substantiated by the similarities in kinematic 
properties and metallicity distributions between: \\
(1) the FHS and the GCs more metal-poor than [Fe/H] $\sim -1$ (``halo GCs");\\
(2) the thick disk stars and the GCs more metal-rich than [Fe/H] $\sim -1$ 
(``disk GCs");\\
(Zinn 1985, Armandroff 1989, Harris 1998).

A few authors (Cayrel 1986; Smith 1986, 1987; Morgan and Lake 1989) have 
considered
the possibility of GCs self-enrichment by SNe and, through simple and 
qualitative arguments, have shown that this was indeed possible under certain 
conditions.

In a recent paper, Brown et al.\  (1995) have developed a more detailed model 
for the early dynamical evolution and self-enrichment of GCs which supports 
the first part of our scenario.  
They show that the SN explosions of the first generation stars trigger the 
formation of an expanding shell, decelerated
by the surrounding hot ISM, in which second generation stars can form.  They 
also discuss the conditions for a GC to survive this phase of chemical 
self-enrichment.

While the second generation stars are forming, the proto-GC may become
unstable and disruption can occur.
This can happen at any time during the SNe phase.
It is important to notice that, at the time of disruption, the
metallicity is fixed by the rate of mixing of the enriched matter expelled 
by SNe with the pre-existing ISM, i.e., by the ratio of the mass of 
the processed 
material to the mass of the ISM. On the other hand,
the value of   [$\alpha$/Fe] depends on the mass distribution
of the Type~II SNe
which have exploded by the time of disruption.
Of course, only the less massive stars are still visible now.  
They form Pop IIa and they appear somewhere on the slowly 
varying branch, depending on the
time at which disruption of the proto-GC occurred.  
All stars originating from a given proto-GC should have the same [$\alpha$/Fe]
value but stars coming from different proto-GCs will have different values. 
Since the disrupted 
proto-GC has 
not completed its chemical evolution, some of the stars dispersed in the 
halo may have a much lower
metallicity than the lowest ones presently observed in the GCs, independently
of their respective [$\alpha$/Fe].

The GC can also survive the SNe phase. 
When all stars more massive than about 8 $M_\odot$ 
have exploded as SNe,
the $\alpha$ and $r$ elements synthesis stops, leading to a typical value of
[$\alpha$/Fe].  Such a view is
 supported by the analysis of Carney (1996) who finds that GCs do not show
any significant variations in [$\alpha$/Fe] despite wide variations in [Fe/H],
age and kinematics.

According to Fig.~7, our scenario requires this typical value of [$\alpha$/Fe]
to be the maximum value observed in the {\it two branches diagram}, 
which in turn
implies an increase of [$\alpha$/Fe] 
with time in a given proto-GC. The end of the SNe phase must indeed correspond
to the bottom of the vertical branch. Such an increase with time, i.e., 
with decreasing mass of the
progenitor, is in agreement with the theoretical yields computed by 
Woosley \& Weaver (1995) although such results are still rather uncertain 
since other computations (e.g. Thielemann et al. 1993) 
show different behaviours.
Moreover, such an interpretation of Fig.~7 implies a nearly solar [Ti/Fe]
for the most massive SNe, which is also in agreement with some of the models 
computed by Woosley \& Weaver (1995).

\begin{figure}
\begin{center}
\leavevmode 
\epsfxsize= 8.5 cm 
\epsffile{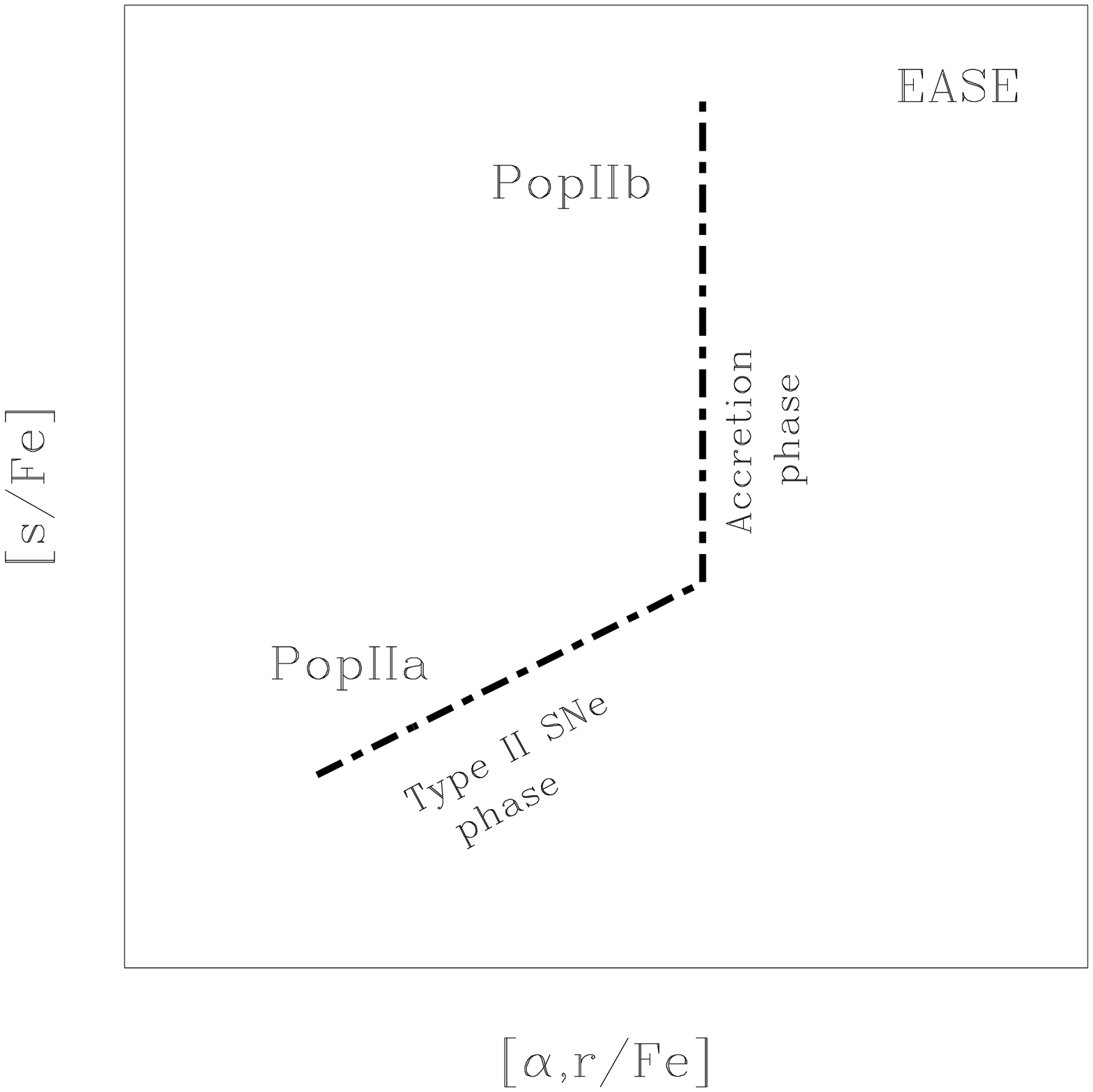}
\vspace*{-5mm}
\caption{EASE scenario.}
\end{center}
\end{figure}

The fact that [Y/Fe] increases with [Ti/Fe] for Pop IIa stars would indicate 
that the ratio of the Y yield to the Fe yield increases with decreasing
mass of the SN progenitor (Fig.~7). On the other hand, the roughly constant 
value obtained for the heavy s elements (Ba, La, Ce) observed in Fig.~9 
suggests that the ratio of the yields (Ba,La,Ce)/Fe does not vary
significantly with the mass of the progenitor.

In a second phase, intermediate mass stars
evolve until they reach the AGB where they enrich their envelope in 
$s$-process elements due to dredge-ups during the thermal pulses.
Through stellar winds or superwinds, those enriched envelopes are ejected 
and they pollute the surrounding less massive stars by accretion.

No new stars would be formed at this stage because the remaining ISM is
probably too diffuse.  During the subsequent evolution of the GC, some of these
surface enriched low mass stars evaporate and form Pop IIb. Those stars can 
also get dispersed in the halo when the GC gets disrupted 
(e.g., when crossing the disk).

For a typical $r$ element like Eu, the behaviour of [Eu/Fe] versus 
[$\alpha$/Fe] (Fig.\ 7)
is completely different, showing a perfect correlation in Pop IIa
stars and an absence of the vertical $s$-process feature,
replaced by a clumping of the points representative of Pop IIb stars at the
constant value of [$\alpha$/Fe] and [$r$/Fe], characteristic of the end of 
the massive
stars outburst.  This shows that, if also produced by lower mass stars, it 
must be in roughly the same proportions as Fe.

Our EASE scenario nicely explains the features observed in the  
{\it two branches diagrams}.  It is schematically displayed in Fig.\  13.

Pop IIa stars mostly originate from disrupted proto-GCs,
their [$\alpha$/Fe] depending on the moment at which
disruption occurs. Pop IIb stars escape later in the 
evolution of the cluster, after the end of the SN phase.

This EASE scenario can also explain the larger metallicity range 
covered by 
the FHS, extending to much lower metallicities than the GCs.  The
very metal-poor stars have escaped from the proto-GCs at a very early stage 
of the outburst phase, when the chemical enrichment of the cloud was still 
very low.

\section*{Acknowledgements}

This work has been supported by contracts ARC 94/99-178 
``Action de Recherche Concert\'ee de la Communaut\'e Fran\c{c}aise de 
Belgique" and P\^ole d'Attraction Interuniversitaire
P4/05 (SSTC, Belgium).  We wish to thank J. Andersen and H. Lindgren 
for providing 
radial velocities for some of the stars and N. Grevesse and J. Sauval for 
some of the atomic data. We also thank N. Grevesse and G. Meylan 
for fruitful discussions.
The Simbad database, operated at CDS, Strasbourg, France, has been used 
in this project.

\end{document}